\documentclass[journal]{IEEEtran}
\usepackage{color,graphicx}
\usepackage[applemac]{inputenc}
\usepackage{epsfig}
\usepackage{latexsym}
\usepackage{amsfonts}
\usepackage{amssymb}
\usepackage{amsmath}
\usepackage{amsbsy}
\usepackage{color,graphicx}
\usepackage{multirow}
\usepackage{xparse}
\usepackage{float}
\usepackage[english]{babel}
\usepackage{url}
\usepackage{cleveref}
\usepackage{subfigure}
\usepackage{ulem}
\usepackage[square, comma, sort&compress,numbers]{natbib}

\graphicspath{{./}{./fig/}}
\usepackage{enumerate}

\newcommand{\argmax}[2]{\smash{\mathop{{\rm argmax}}\limits_{#1}}\, #2 } 

\DeclareTextFontCommand{\emph}{\itshape}

\ifCLASSINFOpdf
\else
\fi
\begin{document}


\title{
Statistics of the MLE and Approximate Upper and Lower Bounds -- Part 1: Application to TOA Estimation 
}


\author{\authorblockN{Achraf~Mallat,~\IEEEmembership{Member,~IEEE,}
				Sinan~Gezici,~\IEEEmembership{Senior Member,~IEEE,}
				Davide~Dardari,~\IEEEmembership{Senior Member,~IEEE,}
				Christophe~Craeye,~\IEEEmembership{Member,~IEEE,}
        and~Luc~Vandendorpe,~\IEEEmembership{Fellow,~IEEE}}
\thanks{Achraf Mallat, Christophe Craeye and Luc Vandendorpe are with the ICTEAM Institute, Universit\'e Catholique de Louvain, Belgium. Email:
\{Achraf.Mallat, Christophe.Craeye, Luc.Vandendorpe\}@uclouvain.be.}
\thanks{Sinan Gezici is with the Department of Electrical and Electronics Engineering, Bilkent University, Ankara 06800, Turkey. Email:
gezici@ee.bilkent.edu.tr.}
\thanks{Davide Dardari is with DEI, CNIT at University of Bologna, Italy. Email: davide.dardari@unibo.it.}
\thanks{This work has been supported in part by the Belgian network IAP Bestcom and the EU network of excellence NEWCOM\#.}
}

\maketitle


\begin{abstract}

In nonlinear deterministic parameter estimation, the maximum likelihood estimator (MLE) is unable to attain the Cramer-Rao lower bound at low and medium signal-to-noise ratios (SNR) due the threshold and ambiguity phenomena.
In order to evaluate the achieved mean-squared-error (MSE) at those SNR levels, we propose new MSE approximations (MSEA) and an approximate upper bound by using the method of interval estimation (MIE).
The mean and the distribution of the MLE are approximated as well.
The MIE consists in splitting the \textit{a priori} domain of the unknown parameter into intervals and computing the statistics of the estimator in each interval.
Also, we derive an approximate lower bound (ALB) based on the Taylor series expansion of noise and an ALB family by employing the binary detection principle.
The accurateness of the proposed MSEAs and the tightness of the derived approximate bounds\footnote{The derived magnitudes are referred as ``bounds" because they are either lower or greater than the MSE, and as ``approximate" because an approximation is performed to obtain them; the terminology ``approximate bound" was previously used by McAulay in \cite{mcaulay3}.} are validated by considering the example of time-of-arrival estimation.

\end{abstract}


\begin{IEEEkeywords}
	Nonlinear estimation, threshold and ambiguity phenomena, maximum likelihood	estimator, mean-squared-error, upper and lowers bounds, time-of-arrival.
\end{IEEEkeywords}

\IEEEpeerreviewmaketitle


\section{Introduction}\label{intro_sec}

\IEEEPARstart{N}{onlinear} estimation of deterministic parameters suffers from the threshold effect \cite{ziv,seidman2,bellini,chow,weiss1,weiss2,zeira1,zeira2,sadler1,sadler2}. This effect means that for a signal-to-noise ratio (SNR) above a given threshold, estimation can achieve the Cramer-Rao lower bound (CRLB), whereas for SNRs lower than that threshold, estimation deteriorates drastically until the estimate becomes uniformly distributed in the \textit{a priori} domain of the unknown parameter.

\smallskip

\begin{figure}
  \centering
  \includegraphics[scale = 0.49]{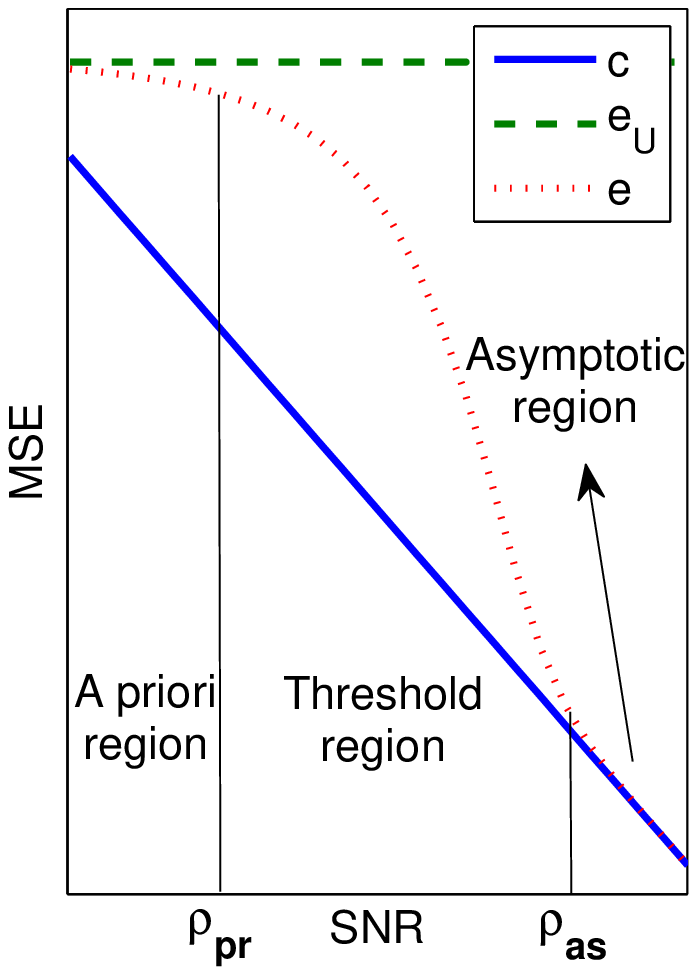}
  \includegraphics[scale = 0.49]{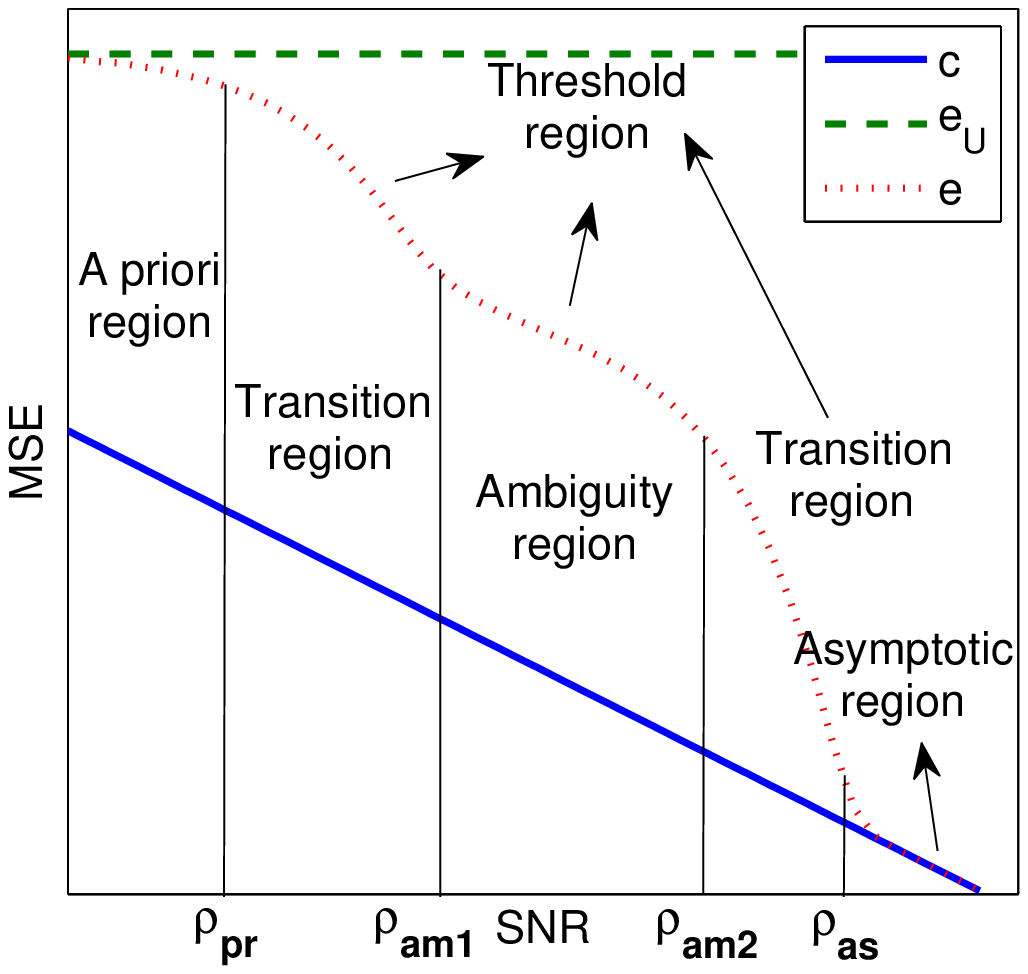}\\
  $\;\;\;\;\;$ (a) 
  $\;\;\;\;\;\;\;\;\;\;\;\;\;\;\;\;\;\;\;\;\;\;\;\;\;\;\;\;\;\;\;\;\;\;\;\;$ 
  (b) $\;\;\;\;\;\;\;\;$
  \caption{SNR regions (a) \textit{A priori}, threshold and asymptotic regions for non-oscillating ACRs (b) \textit{A priori}, ambiguity and asymptotic regions for oscillating ACRs ($c$: CRLB, $e_U$: MSE of uniform distribution in the \textit{a priori} domain, $e$: achievable MSE, $\rho_{pr}$, $\rho_{am1}$, $\rho_{am2}$, $\rho_{as}$: \textit{a priori}, begin-ambiguity, end-ambiguity and asymptotic thresholds).}
  \label{01_regions_pic}
\end{figure}

As depicted in Fig. \ref{01_regions_pic}(a), the SNR axis can be split into three regions according to the achieved mean-squared-error (MSE):
\begin{enumerate}
	\item \textit{A priori} region: Region in which the estimate is uniformly distributed in the \textit{a priori} domain of the unknown parameter (region of low SNRs).
	\item Threshold region: Region of transition between the \textit{a priori} and asymptotic regions (region of medium SNRs).
	\item Asymptotic region: Region in which the CRLB is achieved (region of high SNRs).
\end{enumerate}
In addition, if the autocorrelation (ACR) of the signal carrying the information about the unknown parameter is oscillating, then estimation will be affected by the ambiguity phenomenon \cite[pp. 119]{sinan} and a new region will appear so the SNR axis can be split, as shown Fig. \ref{01_regions_pic}(b), into five regions:
\begin{enumerate}
	\item \textit{A priori} region.
	\item \textit{A priori}-ambiguity transition region.
	\item Ambiguity region.
	\item Ambiguity-asymptotic transition region.
	\item Asymptotic region.
\end{enumerate}
The MSE achieved in the ambiguity region is determined by the envelope of the ACR.
In Figs. \ref{01_regions_pic}(a) and \ref{01_regions_pic}(b), we denote by $\rho_{pr}$, $\rho_{am1}$, $\rho_{am2}$ and $\rho_{as}$ the \textit{a priori}, begin-ambiguity, end-ambiguity and asymptotic thresholds delimiting the different regions.
Note that the CRLB is achieved at high SNRs with asymptotically efficient estimators, such as the maximum likelihood estimator (MLE), only. Otherwise, the estimator achieves its own asymptotic MSE (e.g, MLE with random signals and finite snapshots \cite{Renaux2006,Renaux2007}, Capon algorithm \cite{Richmond2005}).

\smallskip

The exact evaluation of the statistics, in the threshold region, of some estimators such as the MLE has been considered as a prohibitive task.
Many lower bounds (LB) have been derived for both deterministic and Bayesian (when the unknown parameter follows a given \textit{a priori} distribution) parameters in order to be used as benchmarks and to describe the behavior of the MSE in the threshold region \cite{renaux}. 
Some upper bounds (UB) have also been derived like the Seidman UB \cite{seidman1}.
It will suffice to mention here \cite{VanBell2007,renaux}
the Cramer-Rao, Bhattacharyya, Chapman-Robbins, Barankin and Abel deterministic LBs,
the Cramer-Rao, Bhattacharyya, Bobrovsky-MayerWolf-Zakai, Bobrovsky-Zakai, and Weiss-Weinstein Bayesian LBs,
the Ziv-Zakai Bayesian LB (ZZLB) \cite{ziv} with its improved versions: Bellini-Tartara \cite{bellini}, Chazan-Ziv-Zakai \cite{chazan}, Weinstein \cite{weinstein} (approximation of Bellini-Tartara), and Bell-Steinberg-Ephraim-VanTrees \cite{bell} (generalization of Ziv-Zakai and Bellini-Tartara),
and the Reuven-Messer LB \cite{reuven} for problems of simultaneously deterministic and Bayesian parameters.

\smallskip

The CRLB \cite{kay} gives the minimum MSE achievable by an unbiased estimator. However, it is very optimistic for low and moderate SNRs and does not indicate the presence of the threshold and ambiguity regions. The Barankin LB (BLB) \cite{barankin} gives the greatest LB of an unbiased estimator. However, its general form is not easy to compute for most interesting problems. A useful form of this bound, which is much tighter than the CRLB, is derived in \cite{mcaulay1} and generalized to vector cases in \cite{mcaulay2}. 
The bound in \cite{mcaulay1} detects the asymptotic region much below the true one.
Some applications of the BLB can be found in \cite{swerling,seidman2,chow,zeira1,zeira2,knockaert}.

\smallskip

The Bayesian ZZLB family \cite{ziv,chazan,bellini,weinstein,bell} is based on the minimum probability of error of a binary detection problem. 
The ZZLBs are very tight; they detect the ambiguity region roughly and the asymptotic region accurately. 
Some applications of the ZZLBs, discussions and comparison to other bounds can be found in \cite{seidman3,dardari1,sadler1,sadler2,sadler3,sadler4,sinan,dardari2,dardari3,dardari4}.

\smallskip

In \cite[pp. 627-637]{wozencraft}, Wozencraft considered time-of-arrival (TOA) estimation with cardinal sine waveforms and employed the method of interval estimation (MIE) to approximate the MSE of the MLE.
The MIE \cite[pp. 58-62]{VanBell2007} consists in splitting the \textit{a priori} domain of the unknown parameter into intervals and computing the probability that the estimate falls in a given interval, and the estimator mean and variance in each interval.
According to \cite{Van1968,VanBell2007}, the MIE was first used in \cite{Wood1955,Kote1959} before Wozencraft \cite{wozencraft} and others introduced some modifications later.
The approach in \cite{wozencraft} is imitated in \cite{Van1968,rife,Najjar2005,VanBell2007} for frequency estimation and in \cite{Boyer2004} for angle-of-arrival (AOA) estimation. The ACRs in \cite{wozencraft,Van1968,rife,Najjar2005,VanBell2007,Boyer2004,Richmond2005} have the special shape of a cardinal sine (oscillating baseband with the mainlobe twice wider than the sidelobes); this limitation makes their approach inapplicable on other shapes.
In \cite{mcaulay3}, McAulay considered TOA estimation with carrier-modulated pulses (oscillating passband ACRs) and used the MIE to derive an approximate UB (AUB); the approach of McAulay can be applied to any oscillating ACR. 
Indeed, it is followed (independently apparently) in \cite{athley,Richmond2005,Richmond2006} for AOA estimation and in \cite{Najjar2005} (for frequency estimation as mentioned above) where it is compared to Wozencraft's approach. 
The ACR considered in \cite{athley,Richmond2006} has an arbitrary oscillating baseband shape (due to the use of non-regular arrays), meaning that it looks like a cardinal sine but with some strong sidelobes arbitrarily located.
The MSEAs based on Wozencraft's approach are very accurate and the AUBs using McAulay's approach are very tight in the asymptotic and threshold regions. Both approaches can be used to determine accurately the asymptotic region.
Various estimators are considered in the aforecited references.
More technical details about the MIE are given in Sec. \ref{MIE_sec}.

\smallskip

We consider the estimation of a scalar deterministic parameter.
We employ the MIE to propose new approximations (rather than AUBs) of the MSE achieved by the MLE, which are highly accurate, and a very tight AUB. 
The MLE mean and probability density function (PDF) are approximated as well. 
More details about our contributions with regards to the MIE are given in Secs. \ref{MIE_sec} and \ref{upperapprox_sec}.
We derive an approximate LB (ALB) tighter than the CRLB based on the second order Taylor series expansion of noise.
Also, we utilize the binary detection principle to derive some ALBs; the obtained bounds are very tight.
The theoretical results presented in this paper are applicable to any estimation problem satisfying the system model introduced in Sec. \ref{model_sec}.
In order to illustrate the accurateness of the proposed MSEAs and the tightness of the derived bounds, we consider the example of TOA estimation with baseband and passband pulses.

\smallskip

The materials presented in this paper compose the first part of our work divided in two parts \cite{part1,part2}.

\smallskip

The rest of the paper is organized as follows.
In Sec. \ref{model_sec} we introduce our system model. 
In Sec. \ref{thr_amb_sec} we describe the threshold and ambiguity phenomena.
In Sec. \ref{MIE_sec} we deal with the MIE.
In Sec. \ref{upperapprox_sec} we propose an AUB and an MSEA.
In Sec. \ref{lower_sec} we derive some ALBs.
In Sec. \ref{num_sec} we consider the example of TOA estimation and discuss the obtained numerical results.


\section{System model}\label{model_sec}

In this section we consider the general estimation problem of a deterministic scalar parameter (Sec. \ref{general_model_sec}) and the particular case of TOA estimation (Sec. \ref{toa_model_sec}).


\subsection{Deterministic scalar parameter estimation}\label{general_model_sec}

Let $\Theta$ be a deterministic unknown parameter with $D_{\Theta}=[\Theta_1,\Theta_2]$ denoting its \textit{a priori} domain.
We can write the $i$th, $(i=1,\cdots,I)$ observation as:
\begin{equation}\label{r_eq}
	r_i(t) = \alpha s_i(t;\Theta) + \tilde{w}_i(t)
\end{equation}
where $s_i(t;\Theta)$ is the $i$th useful signal carrying the information on $\Theta$, $\alpha$ is a known positive gain, and $\tilde{w}_i(t)$ is an additive white Gaussian noise (AWGN) with two-sided power spectral density (PSD) of $\frac{N_0}{2}$; $\tilde{w}_1(t),\cdots,\tilde{w}_I(t)$ are independent.

\smallskip

Denote by $E_x(\theta)=\sum_{i=1}^{I}\int_{-\infty}^{+\infty}x_i^2(t;\theta)dt$ the sum of the energies of $x_1(t;\theta),\cdots,x_I(t;\theta)$, 
by $\dot{x}$ and $\ddot{x}$ the first and second derivatives of $x$ w.r.t. $\theta$, 
and by $\mathbb{E}$, $\Re$ and $\mathbb{P}$ the expectation, real part and probability operators respectively.
From (\ref{r_eq}) we can write the log-likelihood function of $\Theta$ as:
%
%
\begin{align}
	\Lambda(\theta) = -\frac{1}{N_0}\left[E_r+\alpha^2E_s(\theta)-2\alpha X_{s,r}(\theta)\right] \label{loglikeli2_eq}
\end{align}
where $\theta \in D_{\Theta}$ denotes a variable associated with $\Theta$, and
\begin{equation} \label{crosscorr_eq} 
	X_{s,r}(\theta) = \sum_{i=1}^{I}\int_{-\infty}^{+\infty}s_i(t;\theta)r_i(t)dt = \alpha R_s(\theta,\Theta) + w(\theta)
\end{equation}
is the crosscorrelation (CCR) with respect to (w.r.t.) $\theta$, with
\begin{equation}\label{autocorr_eq}
	R_s(\theta,\theta') = \sum_{i=1}^{I}\int_{-\infty}^{+\infty}s_i(t;\theta)s_i(t;\theta')dt
\end{equation}
denoting the ACR w.r.t. $(\theta,\theta')$ and
\begin{equation}\label{noise_eq}
	w(\theta) = \sum_{i=1}^{I}\int_{-\infty}^{+\infty}s_i(t;\theta)\tilde{w}_i(t)dt
\end{equation}
being a colored zero-mean Gaussian noise of covariance 
\begin{equation} \label{covariance_eq}
	C_{w}(\theta,\theta') = \sum_{i=1}^{I}\mathbb{E}\left\{w_i(\theta)w_i(\theta')\right\} = \frac{N_0}{2}R_s(\theta,\theta').
\end{equation}


\subsubsection{MLE, CRLB and envelope CRLB}

By assuming $E_s(\theta)=E_s$ in (\ref{loglikeli2_eq}), that is, $E_s(\theta)$ is independent of $\theta$, we can respectively write the MLE and the CRLB of $\Theta$ as \cite[pp. 39]{kay}:
\begin{eqnarray}
	\hat{\Theta} &=& \argmax{\theta \in D_{\Theta}}{X_{s,r}(\theta)} \label{mle_eq}\\[0.25 cm]
	c(\Theta) &=& \frac{-1}{\mathbb{E}\{\ddot{\Lambda}(\theta)|_{\theta=\Theta}\}}	
				= \frac{-N_0/2}{\alpha^2\ddot{R}_s(\Theta,\Theta)}
				= \frac{1}{\rho\beta_s^2(\Theta)} \label{crlb_eq}
\end{eqnarray}
where 
\begin{eqnarray}
	\rho &=& \frac{\alpha^2 E_s}{N_0/2} \label{snr_eq}\\
	\beta_s^2(\Theta) &=& -\frac{\ddot{R}_s(\Theta,\Theta)}{E_s} \label{mqbw_eq}
\end{eqnarray}
denote the SNR and the normalized curvature of $R_s(\theta,\Theta)$ at $\theta=\Theta$ respectively.
Unlike $E_s(\Theta)$, $\ddot{R}_s(\Theta,\Theta)$ may depend on $\Theta$ (e.g, AOA estimation \cite{mallat1}).
The CRLB in (\ref{crlb_eq}) is inversely proportional to the curvature of the ACR at $\theta=\Theta$. Sometimes $R_s(\theta,\Theta)$ is oscillating w.r.t. $\theta$. Then, if the SNR is sufficiently high (resp. relatively low) the maximum of the CCR in (\ref{crosscorr_eq}) will fall around the global maximum (resp. the local maxima) of $R_s(\theta,\Theta)$ and the MLE in (\ref{mle_eq}) will (resp. will not) achieve the CRLB.
We will see in Sec. \ref{num_sec} that the MSE achieved at medium SNRs is inversely proportional to the curvature of the envelope of the ACR instead of the curvature of the ACR itself. 
To characterize this phenomenon known as ``ambiguity" \cite{Skolnik} we will define below the envelope CRLB (ECRLB).

\smallskip

Denote by $f$ the frequency\footnote{E.g, $f$ is in seconds (resp. Hz) for frequency (resp. TOA) estimation.} relative to $\theta$ and define the Fourier transform (FT), the mean frequency and the complex envelope w.r.t. $f_c(\Theta)$ of $R_s(\theta,\Theta)$ respectively by
\begin{eqnarray}
	\mathcal{F}_{R_s}(f) &=& \int_{\Theta_1}^{\Theta_2}R_s(\theta,\Theta)e^{-j2\pi f(\theta-\Theta)}d\theta \label{ft_eq}\\
	f_c(\Theta) &=& \frac{\int_0^{+\infty}f \Re\{\mathcal{F}_{R_s}(f)\}df}{\int_0^{+\infty}\Re\{\mathcal{F}_{R_s}(f)\}df} \label{phic_eq}\\
	R_s(\theta,\Theta) &=& \Re\left\{e^{j2\pi(\theta-\Theta)f_c(\Theta)}e_{R_s}(\theta,\Theta)\right\}. \label{cenv_eq}
\end{eqnarray}
In Appendix \ref{curvature_app} we show that: 
\begin{equation}\label{curvature_eq}
	-\ddot{R}_s(\Theta,\Theta) = -\Re\{\ddot{e}_{R_s}(\Theta,\Theta)\} + 4\pi^2f_c^2(\Theta)E_s.
\end{equation}
Now, we define the ECRLB as:
\begin{equation} \label{envcrlb_eq}
	c_e(\Theta) = -\frac{N_0/2}{\alpha^2\Re\{\ddot{e}_{R_s}(\Theta,\Theta)\}}
				= \frac{1}{\rho\beta_e^2(\Theta)}
\end{equation}
where
\begin{eqnarray}
	\beta_e^2(\Theta) = -\frac{\Re\left\{\ddot{e}_{R_s}(\Theta,\Theta)\right\}}{E_s} \label{envmqbw_eq}
\end{eqnarray}
denotes the normalized curvature of $e_{R_s}(\theta,\Theta)$ at $\theta=\Theta$.
From (\ref{mqbw_eq}), (\ref{curvature_eq}) and (\ref{envmqbw_eq}), we have:
%
\begin{equation} \label{mqbwemqbw_eq}
	\beta_s^2(\Theta) = \beta_e^2(\Theta)+4\pi^2f_c^2(\Theta).
\end{equation}	


\subsubsection{BLB}

The BLB can be written as \cite{mcaulay1}:
\begin{equation}\label{barankin_eq}
	c_{B} = (\underline{\Theta}-\Theta)^TD^{-1}(\underline{\Theta}-\Theta)
\end{equation}
where
\begin{eqnarray}
	\underline{\Theta} &=& (\theta_{n_1}\cdots\theta_{-1} \;\; 1+\Theta \;\; \theta_1\cdots\theta_{n_N})^T \nonumber\\
	D &=& (d_{i,j})|_{i,j=n_1,\cdots,n_N} \nonumber
\end{eqnarray}
with $\theta_{n_1},\cdots,\theta_{n_N}$ ($n_1\leq0$, $n_N\geq0$, $\theta_0=\Theta$) denoting $N$ testpoints in the \textit{a priori} domain of $\Theta$, and\footnote{We can show that $E_{\dot{s}}(\theta)=-\ddot{R}_s(\theta,\Theta)$ if $E_s(\theta)$ is independent from $\theta$.}
\begin{align}
	d_{0,0} &= \textstyle \frac{\alpha^2E_{\dot{s}}(\Theta)}{N_0/2} = \frac{1}{c(\Theta)} \nonumber\\
	d_{0,i\neq0} &= d_{i,0} = \textstyle \frac{\alpha^2}{N_0/2}[\dot{R}_s(\Theta,\theta_i)-\dot{R}_s(\Theta,\Theta)] \nonumber\\
	d_{i\neq0,j\neq0} &= \textstyle \frac{\alpha^2}{N_0/2}[R_s(\theta_i,\theta_j)-R_s(\theta_i,\Theta)-R_s(\theta_j,\Theta)+E_s]. \nonumber
\end{align}

\smallskip


\subsubsection{Maximum MSE}

The maximum MSE
\begin{eqnarray}
	e_U = \sigma^2_U+(\Theta-\mu_U)^2 \label{mseU_eq}
\end{eqnarray}
with $\mu_U=\frac{\Theta_1+\Theta_2}{2}$ and $\sigma^2_U=\frac{(\Theta_2-\Theta_1)^2}{12}$ is achieved when the estimator becomes uniformly distributed in $D_{\Theta}$ \cite{dardari1,dardari3}.

\smallskip

The system model considered in this subsection is satisfied for various estimation problems such as TOA, AOA, phase, frequency and velocity estimation. 
Therefore, the theoretical results presented in this paper are valid for the different mentioned parameters. 
TOA is just considered as an example to validate the accurateness and the tightness of our MSEAs and upper and lowers bounds. 


\subsection{Example: TOA estimation}\label{toa_model_sec}

\begin{figure*}[ht]
	\begin{center}
		\subfigure[]{\includegraphics[width=5.9cm]{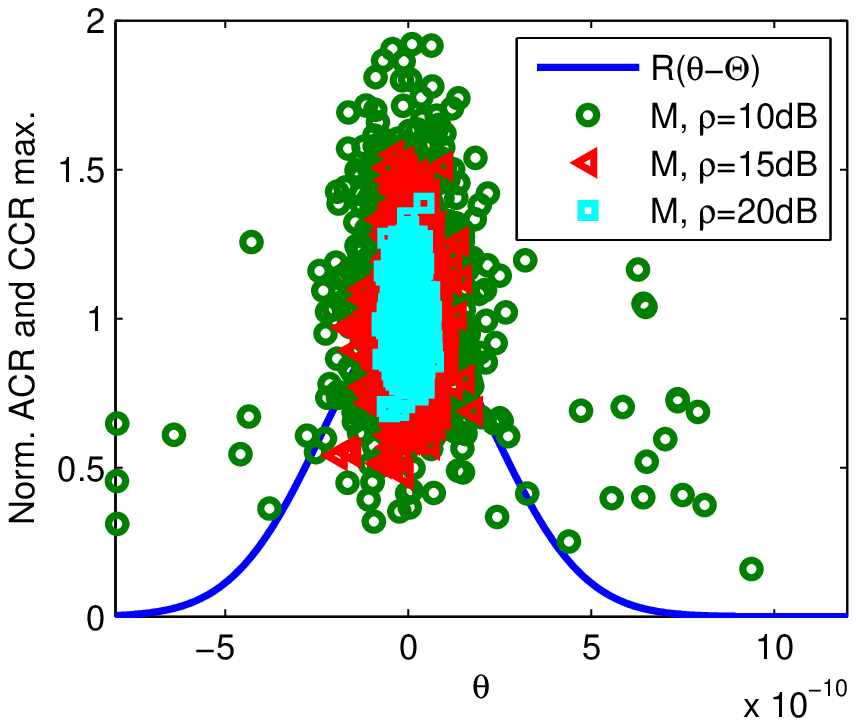} \label{03_mlesamplesFc0_pic}} \hspace{-0.035\textwidth}
		\quad\subfigure[]{\includegraphics[width=5.9cm]{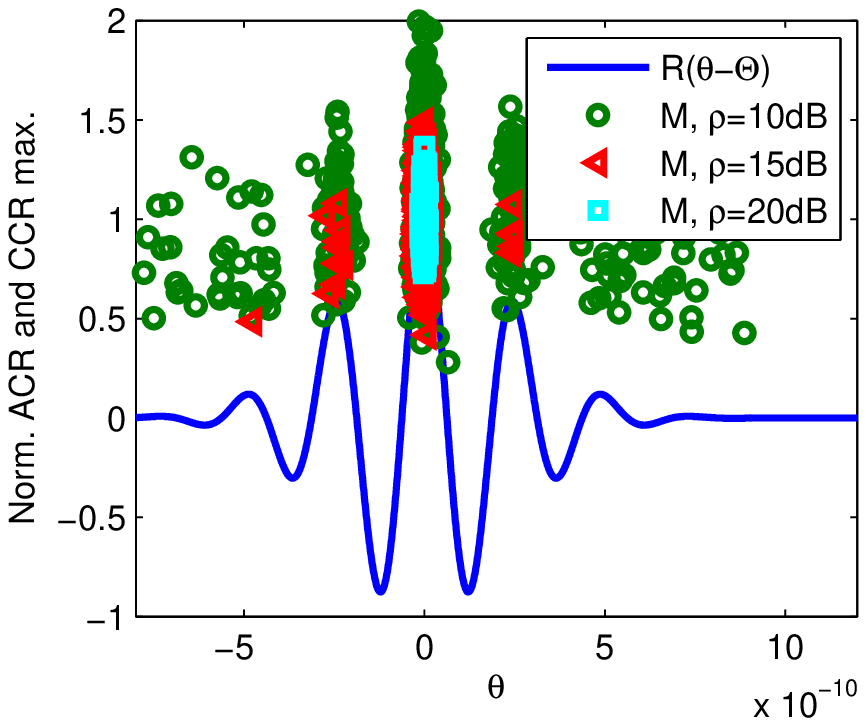} \label{03_mlesamplesFc4_pic}} \hspace{-0.035\textwidth}
		\quad\subfigure[]{\includegraphics[width=5.9cm]{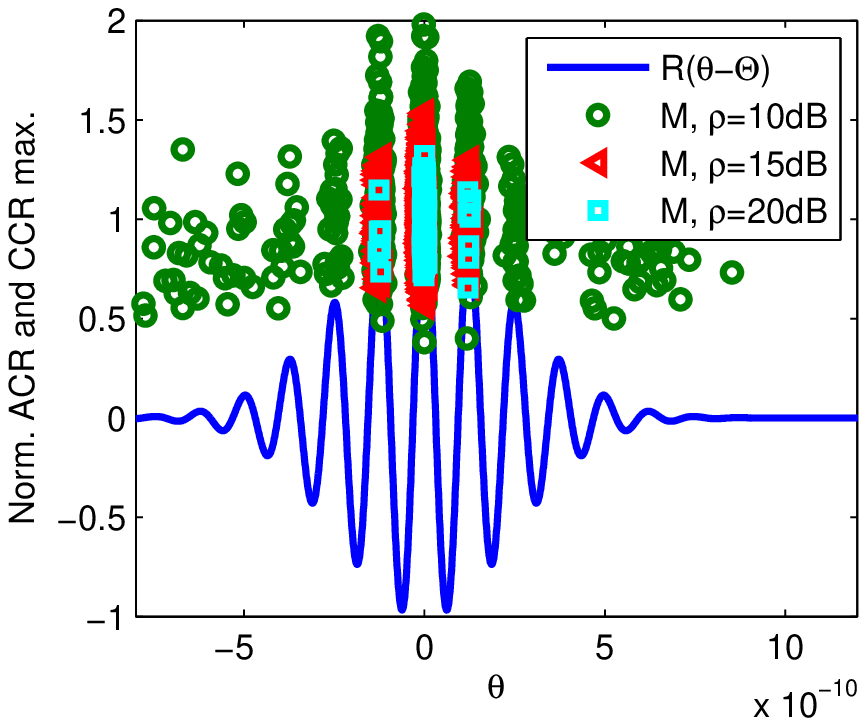} \label{03_mlesamplesFc8_pic}}
	\caption{Normalized ACR $R(\theta-\Theta)$ and 1000 realizations of $M[\hat{\Theta},X(\hat{\Theta})]$ per SNR ($\rho=10$, 15 and 20 dB); Gaussian pulse modulated by $f_c$, $\Theta=0$ ns, $T_w=0.6$ ns, $D_{\Theta}=[-1.5,1.5]T_w$ (a) $f_c=0$ GHz (b) $f_c=4$ GHz (c) $f_c=8$ GHz.}
	\end{center}
\end{figure*}

With TOA estimation based on one observation ($I=1$), $s_1(t;\Theta)$ in (\ref{r_eq}) becomes $s_1(t;\Theta) = s(t-\Theta)$ where $s(t)$ denotes the transmitted signal and $\Theta$ represents the delay introduced by the channel.
Accordingly, we can write the ACR in (\ref{autocorr_eq}) as $R_s(\theta,\theta')=R_s(\theta-\theta')$ where $R_s(\theta) = \int_{-\infty}^{+\infty}s(t+\theta)s(t)dt$, and the CCR in (\ref{crosscorr_eq}) as:
\begin{eqnarray}
	X_{s,r}(\theta) = \alpha R_s(\theta-\Theta) + w(\theta). \label{crosscorr_toa_eq}
\end{eqnarray}
The CRLB $c(\Theta)$ in (\ref{crlb_eq}), ECRLB $c_e(\Theta)$ in (\ref{envcrlb_eq}), mean frequency $f_c(\Theta)$ in (\ref{phic_eq}), normalized curvatures $\beta_s^2(\Theta)$ in (\ref{mqbw_eq}) and $\beta_e^2(\Theta)$ in (\ref{envmqbw_eq}) become now all independent of $\Theta$. 
Furthermore, $\beta_s^2$ and $\beta_e^2$ denote now the mean quadratic bandwidth (MQBW) and the envelope MQBW (EMQBW) of $s(t)$ respectively.

\smallskip

The CRLB in (\ref{crlb_eq}) is much smaller than the ECRLB in (\ref{envcrlb_eq}) because the MQBW in (\ref{mqbwemqbw_eq}) is much larger than the EMQBW in (\ref{envmqbw_eq}). In fact, for a signal occupying the whole band from 3.1 to 10.6 GHz\footnote{
The ultra wideband (UWB) spectrum authorized for unlicensed use by the US federal commission of communications in May 2002 \cite{fcc}.
} ($f_c=6.85$ GHz, bandwidth $B=7.5$ GHz), we obtain $\beta_e^2=\frac{\pi^2B^2}{3}\approx185$ GHz$^2$, $4\pi^2f_c^2\approx10\beta_e^2$, $\beta_s^2\approx11\beta_e^2$ and $c\approx\frac{c_e}{11}$. Therefore, the estimation performance seriously deteriorates at relatively low SNRs when the ECRLB is achieved instead of the CRLB due to ambiguity. 


\section{Threshold and ambiguity phenomena} \label{thr_amb_sec}

In this section we explain the physical origin of the threshold and ambiguity phenomena by considering TOA estimation with UWB pulses\footnote{
We chose UWB pulses because they can achieve the CRLB at relatively low SNRs thanks to their relatively high fractional bandwidth (bandwidth to central frequency ratio).
} as an example. 
The transmitted signal
\begin{eqnarray}
	s(t) = 2sqrt{\frac{E_s}{T_w}}e^{-2\pi\frac{t^2}{T_w^2}}\cos(2\pi f_ct) \label{pulse_eq}
\end{eqnarray}
is a Gaussian pulse of width $T_w$ modulated by a carrier $f_c$.
We consider three values of $f_c$ ($f_c=0$, 4 and 8 GHz) and three values of the SNR ($\rho=10$, 15 and 20 dB) per considered $f_c$. We take $\Theta=0$, $T_w=0.6$ ns, and $D_{\Theta}=[-1.5,1.5]T_w$. 

\smallskip

In Figs. \ref{03_mlesamplesFc0_pic}--\ref{03_mlesamplesFc8_pic} we show the normalized ACR 
$R(\theta-\Theta) = \frac{R_s(\theta-\Theta)}{E_s}$
for $f_c=0$ (baseband pulse), 4 and 8 GHz (passband pulses) respectively, and 1000 realizations per SNR of the maximum 
$M[\hat{\Theta},X(\hat{\Theta})]$
of the normalized CCR $X(\theta)=\frac{X_{s,r}(\theta)}{\alpha E_s}$. 
Denote by $N_n$, ($n=n_1,\cdots,n_N$), ($N$ is the number of local maxima in $D_{\Theta}$), ($n_1<0$, $n_N>0$), ($n=0$ corresponds to the global maximum) the number of samples of $M$ falling around the $n$th local maximum (i.e. between the two local minima adjacent to that maximum) of $R(\theta-\Theta)$.
In Table \ref{03_mlesamples_tab}, we show w.r.t. $f_c$ and $\rho$ the number of samples falling around the maxima number 0 and 1, the CRLB square root (SQRT) $\sqrt{c}$ of $\Theta$, the root MSE (RMSE) $\sqrt{e_S}$ obtained by simulation and the RMSE to CRLB SQRT ratio $\sqrt{\frac{e_S}{c}}$.

\begin{table}
	\begin{center}
		$\begin{array}{|c|c||c|c|c|c|c|}
			\hline
			f_c   &  \rho  &  \sqrt{c}  &  \sqrt{e_S}  &  \sqrt{\frac{e_S}{c}}  &   N_{0}  &  N_1 \\
			\hline\hline
			0 & \begin{array}{c} 10\\15\\20 \end{array} 
			& \begin{array}{c} 76\\43\\24 \end{array} & \begin{array}{c} 123\\46\\24 \end{array} & \begin{array}{c} 1.61\\1.10\\1.01 \end{array} 
			& \begin{array}{c} 1000\\1000\\1000 \end{array} & \begin{array}{c} 0\\0\\0 \end{array} \\
			\hline
			4 & \begin{array}{c} 10\\15\\20 \end{array} 
			& \begin{array}{c} 12\\7\\4 \end{array} & \begin{array}{c} 196\\31\\4 \end{array} & \begin{array}{c} 15.81\\4.47\\1.01 \end{array} 
			& \begin{array}{c} 773\\985\\1000 \end{array} & \begin{array}{c} 59\\8\\0 \end{array} \\
		  \hline
			8 & \begin{array}{c} 10\\15\\20 \end{array} 
			& \begin{array}{c} 6.3\\3.5\\2 \end{array} & \begin{array}{c} 198\\50\\14 \end{array} & \begin{array}{c} 31.56\\14.35\\7.14 \end{array} 
			& \begin{array}{c} 481\\838\\987 \end{array} & \begin{array}{c} 199\\75\\7 \end{array} \\
		  \hline
		\end{array}$
	\end{center}
		\caption{CRLB SQRT $\sqrt{c}$ (ps), simulated RMSE $\sqrt{e_S}$ (ps), RMSE to CRLB SQRT ratio $\sqrt{\frac{e_S}{c}}$, and number ($N_0$, $N_1$) of the $M$ samples falling around the maxima number 0 and 1, for $f_c=0$, 4 and 8 GHz, and $\rho=10$, 15 and 20 dB.}
		\label{03_mlesamples_tab}
\end{table}

\smallskip

Consider first the baseband pulse. We can see in Fig. \ref{03_mlesamplesFc0_pic} that the samples of $M$ are very close to the maximum of $R(\theta-\Theta)$ for $\rho=20$ dB, and they start to spread progressively along $R(\theta-\Theta)$ for $\rho=15$ and 10 dB. 
Table \ref{03_mlesamples_tab} shows that the CRLB is approximately achieved for $\rho=20$ and 15 dB, but not for $\rho=10$ dB.
Based on this observation, we can describe the threshold phenomenon as follows. For sufficiently high SNRs (resp. relatively low SNRs), the maximum of the CCR falls in the vicinity of the maximum of the ACR (resp. spreads along the ACR) so the CRLB is (resp. is not) achieved.

\smallskip

Consider now the pulse with $f_c=4$ GHz. Fig. \ref{03_mlesamplesFc4_pic} and Table \ref{03_mlesamples_tab} show that for $\rho=20$ dB all the samples of $M$ fall around the global maximum of $R(\theta-\Theta)$ and the CRLB is achieved, whereas for $\rho=15$ and 10 dB they spread along the local maxima of $R(\theta-\Theta)$ and the achieved MSE is much larger than the CRLB.
Based on this observation, we can describe the ambiguity phenomenon as follows. For sufficiently high SNRs (resp. relatively low SNRs) the noise component $w(t)$ in the CCR $X_{s,r}(\theta)$ in (\ref{crosscorr_toa_eq}) is not (resp. is) sufficiently high to fill the gap between the global maximum and the local maxima of the ACR. 
Consequently, for sufficiently high SNRs (resp. relatively low SNRs) the maximum of the CCR always falls around the global maximum (resp. spreads along the local maxima) of the ACR so the CRLB is (resp. is not) achieved.
Obviously, the ambiguity phenomenon affects the threshold phenomenon because the SNR required to achieve the CRLB depends on the gap between the global and the local maxima.

\smallskip

Let us now examine the RMSE achieved at $\rho=20$ dB for $f_c=4$ and $8$ GHz; it is 3.5 times smaller with $f_c=4$ GHz than with $f_c=8$ GHz whereas the CRLB SQRT is 2 times smaller with the latter.
In fact, the samples of $M$ do not fall all around the global maximum for $f_c=8$ GHz.
This amazing result (observed in \cite{mallat3} from experimental results) exhibits the significant loss in terms of accuracy if the CRLB is not achieved due to ambiguity.
It also shows the necessity to design our system such that the CRLB be attained.


\section{
MIE-based MLE statistics approximation
}\label{MIE_sec}

We have seen in Sec. \ref{thr_amb_sec} that the threshold phenomenon is due to the spreading of the estimates along the ACR.
To characterize this phenomenon we split the \textit{a priori} domain $D_{\Theta}$ into $N$ intervals $D_n=[d_n,d_{n+1})$, $(n=n_1,\cdots,n_N)$, ($n_1\leq0$, $n_N\geq0$) and write the PDF, mean and MSE of $\hat{\Theta}$ as
\begin{align}
	p(\theta) &= \sum_{n=n_1}^{n_N}P_np_n(\theta) \nonumber\\
	\mu &= \int_{\Theta_1}^{\Theta_2}\theta p(\theta)d\theta = \sum_{n=n_1}^{n_N}P_n\mu_n \nonumber\\
	e &= \int_{\Theta_1}^{\Theta_2}(\theta-\Theta)^2p(\theta)d\theta = \sum_{n=n_1}^{n_N}P_n\left[\left(\Theta-\mu_n\right)^2+\sigma^2_n\right] \label{mse_eq}
\end{align}
where
\begin{align}
	P_n &= \mathbb{P}\{\hat{\Theta}\in D_n\} \label{Pn_eq}\\
				&= \mathbb{P}\{\exists \xi \in D_n: X_{s,r}(\xi)>X_{s,r}(\theta), \forall\theta\in\cup_{n'\neq n}D_{n'}\} \nonumber
\end{align}
denotes the interval probability (i.e. probability that $\hat{\Theta}$ falls in $D_n$),
and $p_n(\theta)$, $\mu_n=\mathbb{E}\{\hat{\Theta}_n\}$ and $\sigma_n^2=\mathbb{E}\{(\hat{\Theta}_n-\mu_n)^2\}$ represent, respectively, the PDF, mean and variance of the interval MLE ($\hat{\Theta}$ given $\hat{\Theta}\in D_n$)
\begin{equation} \label{mlen_eq}
	\hat{\Theta}_n=\hat{\Theta}\big|\hat{\Theta}\in D_n.
\end{equation}
Denote by $\theta_n$ a testpoint selected in $D_n$ and let $X_n = X_{s,r}(\theta_n) = \alpha R_n + w_n$ with $R_n=R_s(\theta_n,\Theta)$ and $w_n=w(\theta_n)$.
Using (\ref{crosscorr_eq}), $P_n$ in (\ref{Pn_eq}) can be approximated by
\begin{align}
	\tilde{P}_n &= \mathbb{P}\{X_n>X_{n'}, \forall n'\neq n\} 
				= \int_{-\infty}^{+\infty}dx_n \int_{-\infty}^{x_n}dx_{n_1}\cdots \nonumber\\ 
						& \int_{-\infty}^{x_n}dx_{n-1} \int_{-\infty}^{x_n}dx_{n+1}\cdots\int_{-\infty}^{x_n}p_{X}(x)dx_{n_N} \label{Pnapp_eq}
\end{align}
where
\begin{eqnarray}
	p_{X}(x) = \frac{1}{(2\pi)^{\frac{N}{2}}|C_X|^{\frac{1}{2}}}e^{-\frac{(x-\mu_X)C_X^{-1}(x-\mu_X)^T}{2}} \nonumber
\end{eqnarray}
represents the PDF of $X=(X_{n_1}\cdots X_{n_N})^T$ with 
$\mu_X = (\mu_{X_{n_1}}\cdots\mu_{X_{n_N}})^T = \alpha(R_{n_1}\cdots R_{n_N})^T$ being its mean and 
$C_X = \frac{N_0}{2}\left[R_s(\theta_n,\theta_{n'})\right]_{n,n'=n_1,\cdots,n_N}$ its covariance matrix.

\smallskip

The accuracy of the approximation in (\ref{Pnapp_eq}) depends on the choice of the intervals and the testpoints. 
For an oscillating ACR we consider an interval around each local maximum and choose the abscissa of the local maximum as a testpoint, whereas for a non-oscillating ACR we split $D_{\Theta}$ into equal intervals and choose the center $\theta_n=\frac{d_n+d_{n+1}}{2}$ of each interval as a testpoint.
For both oscillating and non-oscillating ACRs, $D_0$ contains the global maximum and $\theta_0$ is equal to $\Theta$.

\smallskip

The testpoints are chosen as the roots of the ACR (except for $\theta_0=\Theta$) in \cite{wozencraft,Van1968,rife,Boyer2004,Najjar2005,VanBell2007}, as the local extrema abscissa in \cite{mcaulay3}, and as the local maxima abscissa in \cite{Najjar2005,athley,Richmond2005,Richmond2006}.


\subsection{Computation of the interval probability} \label{subproba_sec}

We consider here the computation of the approximate interval probability $\tilde{P}_n$ in (\ref{Pnapp_eq}).

\smallskip


\subsubsection{Numerical approximation}

To the best of our knowledge there is no closed form expression for the integral in (\ref{Pnapp_eq}) for correlated $X_n$. However, it can be computed numerically using for example the MATLAB function QSCMVNV (written by Genz based on \cite{genz1,genz2,genz3,genz4}) that computes the multivariate normal probability with integration region specified by a set of linear inequalities in the form $b_1<B(X-\mu_X)<b_2$. Using QSCMVNV, $\tilde{P}_n$ can be approximated by:
\begin{equation}
	P^{(1)}_n = \text{QSCMVNV}(N_p,C_X,b_1,B,b_2) \label{proba1_eq}
\end{equation}
where $N_p$ is the number of points used by the algorithm (e.g, $N_p=3000$), 
$b_1=(-\infty\cdots-\infty)^T$ and 
$b_2=\mu_{X_n}-(\mu_{X_{n_1}}\cdots\mu_{X_{n-1}}\mu_{X_{n+1}}\cdots\mu_{X_{n_N}})^T$ two $(N-1)$-column vectors, and
$B=\left(\begin{array}{c|c|c}\begin{array}{c} B_1\\B_2 \end{array} & B_3 & \begin{array}{c} B_4\\B_5 \end{array} \end{array}\right)$ an $(N-1)\times N$ matrix
with $B_1=I(n-n_1)$, $B_2=\text{zeros}(N+n_1-n-1,n-n_1)$, $B_3=-\text{ones}(N-1,1)$, $B_4=\text{zeros}(N-n_N+n-1,n_N-n)$ and $B_5=I(n_N-n)$\footnote{We denote by $I(k)$ the identity matrix of rank $k$, and $\text{zeros}(k_1,k_2)$ and $\text{ones}(k_1,k_2)$ the zero and one matrices of dimension $k_1\times k_2$.}.

\smallskip


\subsubsection{Analytic approximation}

Denote by $Q(y)=\frac{1}{\sqrt{2\pi}}\int_{y}^\infty e^{-\frac{\xi^2}{2}}d\xi$ the Q function. As $\mathbb{P}\{A_1\cap A_2\}\leq \mathbb{P}\{A_1\}$, we can upper bound $\tilde{P}_n$ in (\ref{Pnapp_eq}) by:
\begin{equation} \label{proba2_eq}
	P^{(2)}_n = \left\{\begin{array}{ll}
				P(\theta_0,\theta_1) & n=0 \\
				P(\theta_n,\theta_0) & n\neq0 \end{array} \right.
\end{equation}
where
\begin{align}
	P(\theta,\theta') &= \mathbb{P}\{X_{s,r}(\theta)>X_{s,r}(\theta')\} \nonumber\\
				&= Q\left(\sqrt{\frac{\rho}{2}}\frac{R(\theta',\Theta)-R(\theta,\Theta)}{\sqrt{1-R(\theta,\theta')}}\right) \label{Q_eq}
\end{align}
with $R(\theta,\Theta) = \frac{R_s(\theta,\Theta)}{E_s}$ denoting the normalized ACR.
$P(\theta,\theta')$ is obtained (\ref{Q_eq}) from (\ref{crosscorr_eq}) and (\ref{covariance_eq}) by noticing that
$X_{s,r}(\theta)-X_{s,r}(\theta')\sim\mathcal{N}(\alpha[R_s(\theta,\Theta)-R_s(\theta',\Theta)],N_0[E_s-R_s(\theta,\theta')])$\footnote{$\mathcal{N}(m,v)$ stands for the normal distribution of mean $m$ and variance $v$.}.
If $N$ approaches infinity, then both $\sum_{n=n_1}^{n_N}P^{(2)}_n$ and the MSEA in (\ref{mse_eq}) will approach infinity. 

\begin{figure}[t]
  \centering
  \includegraphics[width=8cm]{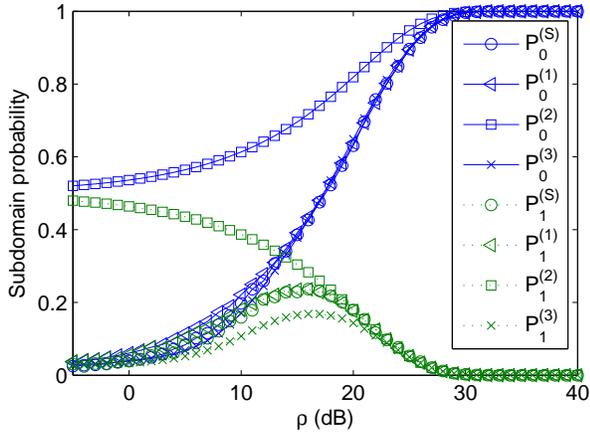}
  \caption{Simulated interval probability $P^{(S)}_n$, the approximations $P^{(1)}_n$ and $P^{(3)}_n$, and the AUB $P^{(2)}_n$ for $n=0,1$ w.r.t. the SNR.}
  \label{06_Pn_snr_pic}
\end{figure}

\smallskip

Using \eqref{proba2_eq}, we propose the following approximation:
\begin{equation} \label{proba3_eq}
	P^{(3)}_n = \frac{P^{(2)}_n}{\sum_{n=n_1}^{n_N}P^{(2)}_n}.
\end{equation}

In this subsection we have seen that the interval probability $P_n$ in (\ref{Pn_eq}) can be approximated by $P^{(1)}_n$ in (\ref{proba1_eq}) or $P^{(3)}_n$ in (\ref{proba3_eq}), and upper bounded by $P^{(2)}_n$ in (\ref{proba2_eq}).

\smallskip

The UB $P^{(2)}_n$ is adopted in \cite{mcaulay3,Najjar2005,athley,Richmond2005,Richmond2006} with minor modifications; in fact, $\tilde{P}_0$ is approximated by one in \cite{mcaulay3} and by $1-\sum_{n\neq0}P^{(2)}_n$ in \cite{Najjar2005,athley,Richmond2005,Richmond2006}.
%
%
In the special case where $X_{n_1},\cdots,X_{-1},X_{1},\cdots,X_{n_N}$ are independent and identically distributed such as in \cite{wozencraft,Van1968,rife,Boyer2004,Najjar2005,VanBell2007} thanks to the cardinal sine ACR, then $\tilde{P}_n=\frac{\tilde{P}_A}{N-1}$, $\forall n\neq0$, and $\tilde{P}_0=1-\tilde{P}_A$ ($\tilde{P}_A$ is the approximate probability of ambiguity); consequently, the MSEA in \eqref{mse_eq} can be written as the sum of two terms: $e \approx \tilde{P}_Ae_U+\tilde{P}_0c(\Theta)$;
$\tilde{P}_0$ can be calculated by performing one-dimensional integration.
If $X_0\sim\mathcal{N}(\alpha E_s,\frac{N_0}{2}E_s)$ and $X_n\sim\mathcal{N}(0,\frac{N_0}{2}E_s)$, $\forall n\neq0$, like in \cite{wozencraft,Van1968,Najjar2005,VanBell2007} then $P_A$ can be upper bounded using the union bound \cite{wozencraft}.

\smallskip

As an example, to evaluate the accurateness of $P^{(1)}_n$ in (\ref{proba1_eq}) and $P^{(3)}_n$ in (\ref{proba3_eq}) and to compare them to $P^{(2)}_n$ in (\ref{proba2_eq}), we consider the pulse in (\ref{pulse_eq}) with $f_c=6.85$ GHz, $T_w=2$ ns, $\Theta=0$ and $D_{\Theta}=[-2,1.5]T_w$. 
In Fig. \ref{06_Pn_snr_pic} we show for $n=0$ and $1$, the interval probability $P^{(S)}_n$ obtained by simulation based on 10000 trials, $P^{(1)}_n$, $P^{(2)}_n$ and $P^{(3)}_n$, all versus the SNR. 
We can see that $P^{(S)}_n$ converges to $\frac{1}{N}$ at low SNRs for all intervals; however, it converges to $1$ at high SNRs ($P^{S}_0=0.99$ for $\rho\approx30$ dB) for $n=0$ (probability of non-ambiguity) and to 0 for $n\neq0$. Both $P^{(1)}_n$ and $P^{(3)}_n$ are very accurate and closely follow $P^{(S)}_n$.
The UB $P^{(2)}_n$ is not tight at low SNRs; it converges to $0.5$ $\forall n$ instead of $\frac{1}{N}$ due to (\ref{Q_eq}). However, it converges to 1 (resp. 0) for $n=0$ (resp. $n\neq0$) at high SNRs simultaneously with $P^{(S)}_n$ so it can be used to determine accurately the asymptotic region.


\subsection{Statistics of the interval MLE}\label{submle_sec}

We approximate here the statistics of the interval MLE $\hat{\Theta}_n$ in (\ref{mlen_eq}).
We have already mentioned in Sec. \ref{MIE_sec} that for an oscillating (resp. a non-oscillating) ACR we consider an interval around each local maximum (resp. split the \textit{a priori} domain into equal intervals); the global maximum is always contained in $D_0$.
Accordingly, the ACR inside a given interval is either increasing then decreasing 
or monotone (i.e. increasing, decreasing or constant).

\smallskip

As the distribution of $\hat{\Theta}_n$ should follow the shape of the ACR in the considered interval, the interval variance is upper bounded by the variance of uniform distribution in $D_n=[d_n,d_{n+1}]$. Therefore, the interval mean $\mu_n$ and variance $\sigma_n^2$ can be approximated by 
\begin{eqnarray}
	\mu_{n,U} &=& \frac{d_{n}+d_{n+1}}{2} \label{munU_eq}\\
	\sigma^2_{n,U} &=& \frac{(d_{n+1}-d_n)^2}{12}. \label{varnU_eq}
\end{eqnarray}
For intervals with local minima (not considered here), the ACR decreases then increases so $\sigma^2_{n}$ is upper bounded by the variance of a Bernoulli distribution of two equiprobable atoms:
\begin{equation}\label{varnmaxB_eq}
	\sigma^2_{n,\max} = \frac{(d_{n+1}-d_n)^2}{4} > \sigma^2_{n,U}.
\end{equation}
In \cite{mcaulay3}, it is assumed that $\sigma^2_{n}$ is upper bounded by $\sigma^2_{i,U}$ in (\ref{varnU_eq}) even for intervals with local minima.
See \cite{jacobson,dharmadhikari} for further information on the maximum variance.

\smallskip

The CCR $X_{s,r}(\theta)$ in (\ref{crosscorr_eq}) can be approximated inside $D_n$ by its Taylor series expansion about $\theta_n$ limited to second order:
\begin{align}
	X_{s,r}(\theta) &= \alpha R_s(\theta,\Theta) + w(\theta) \nonumber\\
				&\approx (\alpha R_n+w_n) + (\alpha\dot{R}_n+\dot{w}_n)(\theta-\theta_n) \nonumber\\ 
				&+ (\alpha\ddot{R}_n+\ddot{w}_n)\frac{(\theta-\theta_n)^2}{2} \label{crosscorr2nd_eq}
\end{align}
where $\dot{w}_n=\dot{w}(\theta_n)$, $\ddot{w}_n=\ddot{w}(\theta_n)$, $\dot{R}_n=\dot{R}_s(\theta_n,\Theta)$ and $\ddot{R}_n=\ddot{R}_s(\theta_n,\Theta)$. 
Let $\nu_n$ be the correlation coefficient of $\dot{w}_n$ and $\ddot{w}_n$. Then, from (\ref{noise_eq}), we can show that
\begin{eqnarray}
	\dot{w}_n &\sim& \mathcal{N}(0,\sigma^2_{\dot{w}_n}) \label{noise1stderivpdf_eq}\\
	\ddot{w}_n &\sim& \mathcal{N}(0,\sigma^2_{\ddot{w}_n}) \label{noise2ndderivpdf_eq}
\end{eqnarray}
with
\begin{eqnarray}
	\sigma^2_{\dot{w}_n} &=& \frac{N_0}{2}\int_{-\infty}^{+\infty}\dot{s}^2(t;\theta_n)dt = \frac{N_0}{2}E_{\dot{s}}(\theta_n) \label{varnoise1stderiv_eq}\\
	\sigma^2_{\ddot{w}_n} &=& \frac{N_0}{2}\int_{-\infty}^{+\infty}\ddot{s}^2(t;\theta_n)dt = \frac{N_0}{2}E_{\ddot{s}}(\theta_n) \label{varnoise2ndderiv_eq}\\
	\nu_n &=& \frac{\mathbb{E}\{\dot{w}_n\ddot{w}_n\}}{\sigma_{\dot{w}_n}\sigma_{\ddot{w}_n}} 
				= \frac{\int_{-\infty}^{+\infty}\dot{s}(t;\theta_n)\ddot{s}(t;\theta_n)dt}{\sqrt{E_{\dot{s}}(\theta_n)E_{\ddot{s}}(\theta_n)}}. \label{corrcoef_eq}
\end{eqnarray}

Let us first consider an interval with monotone ACR. By neglecting $\ddot{w}_n$ and $\ddot{R}_n$ in (\ref{crosscorr2nd_eq}) (linear approximation), we can approximate the interval MLE by:
\begin{align}
	\hat{\Theta}_n &= \argmax{\theta\in D_n}{\{X_{s,r}(\theta)\}} \nonumber\\[0.15cm]
				& \approx \left\{\begin{array}{cl}
						d_n & \alpha\dot{R}_n+\dot{w}_n<0 \\
						d_{n+1} & \alpha\dot{R}_n+\dot{w}_n>0 \\
						\frac{d_{n,1}+d_{n,2}}{2} & \alpha\dot{R}_n+\dot{w}_n=0. \end{array}\right. \label{condmle2_eq}
\end{align}
As $\mathbb{P}\{\alpha\dot{R}_n+\dot{w}_n=0\}=0$, the latter approximation follows a two atoms Bernoulli distribution with probability, mean and variance given from (\ref{snr_eq}), (\ref{noise1stderivpdf_eq}) and (\ref{varnoise1stderiv_eq}) by:
\begin{eqnarray}
	\mathbb{P}\{d_n\} &=& 1-\mathbb{P}\{d_{n+1}\} = \mathbb{P}\{-\dot{w}_n>\alpha\dot{R}_n\}  \nonumber\\
				&=& Q\Big(\frac{\alpha\dot{R}_n}{\sigma_{\dot{w}_n}}\Big)  = Q\Bigg(\sqrt{\frac{\rho\dot{R}_n^2}{E_sE_{\dot{s}}(\theta_n)}}\Bigg) \label{PnB_eq}\\
	\mu_{n,B} &=& d_n\mathbb{P}\{d_n\}+d_{n+1}\mathbb{P}\{d_{n+1}\} \nonumber\\
	\sigma^2_{n,B} &=& \mathbb{P}\{d_n\}\mathbb{P}\{d_{n+1}\}(d_{n+1}-d_n)^2 \nonumber
\end{eqnarray}
where $\sigma^2_{n,B}$ is upper bounded by $\sigma^2_{n,\max}$ in (\ref{varnmaxB_eq}) and reaches it for $\mathbb{P}\{d_n\}=0.5$; $\mathbb{P}\{d_n\}=0.5$ just means that $\hat{\Theta}_n$ is uniformly distributed in $D_n$ (because $\hat{\Theta}_n$ can fall anywhere inside $D_n$); therefore, $\mu_n$ and $\sigma_n^2$ can be approximated by:
\begin{eqnarray}
	\mu_{n,1,c} &=& \mu_{n,B} \label{mun1c_eq}\\
	\sigma^2_{n,1,c} &=& \min\{\sigma^2_{n,U},\sigma^2_{n,B}\}. \label{varn1c_eq}
\end{eqnarray}
By neglecting $\dot{w}_n$ in (\ref{crosscorr2nd_eq}) and (\ref{condmle2_eq}) (because $\sigma^2_n<<(\Theta-\mu_n)^2$ for $n\neq0$, see (\ref{mse_eq})) we obtain the following approximation:
\begin{eqnarray}
	\mu_{n,2,c} &=& \left\{\begin{array}{cl}
		d_n & \dot{R}_n<0 \\
		d_{n+1} & \dot{R}_n>0 \\
		\frac{d_n+d_{n+1}}{2} & \dot{R}_n=0 \end{array}\right. \label{mun2c_eq}\\
	\sigma^2_{n,2,c} &=& 0. \label{varn2c_eq}
\end{eqnarray}

Consider now an interval with a local maximum. By neglecting $\ddot{w}_n$ in (\ref{crosscorr2nd_eq}), and taking into account that $\dot{R}_n=0$ (local maximum), $\hat{\Theta}_n$ can be approximated by:
\begin{equation} \label{condmle4_eq}
	\hat{\Theta}_n = \argmax{\theta\in D_n}{\{X_{s,r}(\theta)\}} \approx \theta_n -\frac{\dot{w}_n}{\alpha\ddot{R}_n}
\end{equation}
which follows a normal distribution whose PDF, mean and variance can be obtained from (\ref{crlb_eq}), (\ref{noise1stderivpdf_eq}), (\ref{varnoise1stderiv_eq}) and (\ref{condmle4_eq}):
\begin{eqnarray}
	p_{n,N}(\theta) &=& \frac{1}{\sqrt{2\pi}\sigma_{n,N}}e^{-\frac{(\theta-\mu_{n,N})^2}{2\sigma^2_{n,N}}} \label{pdfnN_eq}\\
	\mu_{n,N} &=& \theta_n \label{munN_eq}\\
	\sigma^2_{n,N} &=& \frac{\sigma^2_{\dot{w}_n}}{\alpha^2\ddot{R}_n^2} = \frac{\frac{N_0}{2}E_{\dot{s}}(\theta_n)}{\alpha^2\ddot{R}_n^2} 
				= c\frac{-\ddot{R}_0E_{\dot{s}}(\theta_n)}{\ddot{R}_n^2}. \label{varnN_eq}
\end{eqnarray}
For $n=0$, $\sigma^2_{n,N}$ is equal to the CRLB in (\ref{crlb_eq}) since $-\ddot{R}_0=E_{\dot{s}}(\theta_0)$.
To take into account that $D_n$ is finite, we propose from (\ref{pdfnN_eq}), (\ref{munN_eq}) and (\ref{varnN_eq}) the following approximation:
\begin{eqnarray}
	\mu_{n,1,o} &=& \int_{d_n}^{d_{n+1}}\theta p_{n,1,o}(\theta)d\theta \approx \theta_n \label{mun1o_eq}\\
	\sigma^2_{n,1,o} &=& \int_{d_n}^{d_{n+1}}(\theta-\mu_{n,1,o})^2p_{n,1,o}(\theta)d\theta \nonumber\\
				&\approx& \min\{\sigma^2_{n,N},\sigma^2_{n,U}\} \label{varn1o_eq}
\end{eqnarray}
where $p_{n,1,o}(\theta)=\frac{p_{n,N}(\theta)}{\int_{d_n}^{d_{n+1}}p_{n,N}(\theta)d\theta}$.
By neglecting $w(\theta)$ in (\ref{crosscorr2nd_eq}) and (\ref{condmle4_eq}), we obtain the following approximation:
\begin{eqnarray}
	\mu_{n,2,o} &=& \theta_n \label{mun2o_eq}\\
	\sigma^2_{n,2,o} &=& 0. \label{varn2o_eq}
\end{eqnarray}

\begin{figure}[t]
  \centering
  \includegraphics[width=8cm]{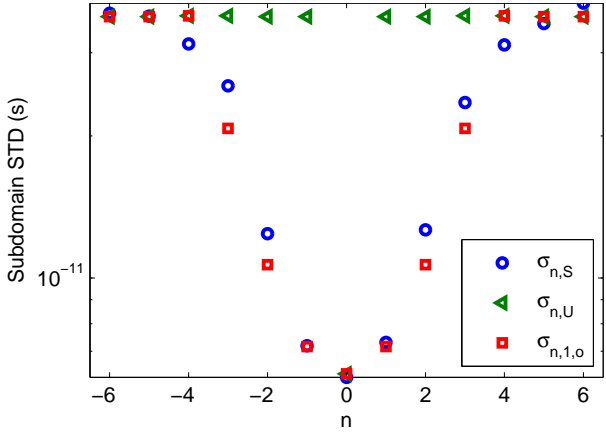}
  \caption{Simulated interval STD $\sigma_{n,S}$ and approximations $\sigma_{n,U}$ and $\sigma_{n,1,o}$ w.r.t. the interval number $n=-6,\cdots,6$ for $\rho=10$ dB.}
  \label{04_varn_pic}
\end{figure}

For both oscillating and non-oscillating ACRs, $D_0$ contains the global maximum.
To guarantee the convergence of the MSEA in (\ref{mse_eq}) to the CRLB, $\mu_0$ and $\sigma^2_0$ should always be approximated using (\ref{mun1o_eq}) and (\ref{varn1o_eq}) by:
\begin{eqnarray}
	\mu_{0,0} &=& \Theta \label{mu0_eq}\\
	\sigma^2_{0,0} &=& \min\{c,\sigma^2_{0,U}\}. \label{var0_eq}
\end{eqnarray}
For TOA estimation, we can write (\ref{PnB_eq}) and (\ref{varnN_eq}) as 
$\mathbb{P}\{d_n\} = Q\left(\sqrt{\rho}\frac{\dot{R}_n}{E_s\beta_s}\right)$ and 
$\sigma^2_{n,N} = c\frac{\ddot{R}_0^2}{\ddot{R}_n^2}$.

\smallskip

We have seen in this subsection that the interval mean and variance can be approximated by
\begin{itemize}
	\item $\mu_{0,0}$ in (\ref{mu0_eq}) and $\sigma^2_{0,0}$ in (\ref{var0_eq}) for $n=0$.
	\item $\mu_{n,U}$ in (\ref{munU_eq}) and $\sigma^2_{n,U}$ in (\ref{varnU_eq}), $\mu_{n,1,c}$ in (\ref{mun1c_eq}) and $\sigma^2_{n,1,c}$ in (\ref{varn1c_eq}), or $\mu_{n,2,c}$ in (\ref{mun2c_eq}) and $\sigma^2_{n,2,c}$ in (\ref{varn2c_eq}) for intervals with monotone ACR.
	\item $\mu_{n,U}$ and $\sigma^2_{n,U}$, $\mu_{n,1,o}$ in (\ref{mun1o_eq}) and $\sigma^2_{n,1,o}$ in (\ref{varn1o_eq}), or $\mu_{n,2,o}$ in (\ref{mun2o_eq}) and $\sigma^2_{n,2,o}$ in (\ref{varn2o_eq}) for intervals with local maxima.
\end{itemize}

\smallskip

In \cite{wozencraft,Van1968,rife,Boyer2004,VanBell2007} (resp. \cite{athley,Richmond2005,Najjar2005,Richmond2006}) $\sigma^2_n$ is approximated by $\sigma^2_{n,U}$ (resp. $\sigma^2_{n,2,o}$). They all approximate $\mu_n$ by $\theta_n$ and $\sigma^2_0$ by the asymptotic MSE (equal to the CRLB if the considered estimator is asymptotically efficient).

\smallskip

To evaluate the accurateness of $\sigma^2_{n,U}$ in (\ref{varnU_eq}) and $\sigma^2_{n,1,o}$ in (\ref{varn1o_eq}), we consider the pulse in (\ref{pulse_eq}) with $f_c=8$ GHz, $T_w=0.6$ ns, $D_{\Theta}=[-1.5,1.5]T_w$ and $\rho=10$ dB. 
In Fig. \ref{04_varn_pic} we show the approximate interval standard deviations (STD) $\sigma_{n,U}$ and $\sigma_{n,1,o}$, and the STD $\sigma_{n,S}$ obtained by simulation based on 50000 trials, w.r.t. the interval number $n=-6,\cdots,6$.
We can see that $\sigma_{n,S}$ is upper bounded by $\sigma_{n,U}$ as expected and that $\sigma_{n,1,o}$ follows $\sigma_{n,S}$ closely. The smallest variance corresponds to $n=0$ because the curvature of $R_s(\theta,\Theta)$ reaches its maximum at $\theta=\Theta$.

\smallskip

Before ending this section, we would like to highlight our contributions regarding the MIE.
We have proposed two approximations for the interval probability when $X_{n_1},\cdots,X_{n_N}$ are correlated.
We have shown in Fig. \ref{06_Pn_snr_pic} how our approximations are accurate. To the best of our knowledge all previous authors adopt the McAulay probability UB (except for the case where $X_{n_1},\cdots,X_{n_N}$ are independent thanks to the cardinal sine ACR).
We have proposed two new approximations for the interval mean and variance, one for intervals with monotone ACRs and one for intervals with local maxima. We have seen in Fig. \ref{04_varn_pic} how our approximations are accurate. To the best of our knowledge all previous authors either upper bound the interval variance or neglect it.
Thanks to the proposed probability approximations our MSEAs (e.g, $e_{1,1,c}$ in Fig. \ref{06_varBBapprox_snr}) are highly accurate and outperform the MSE UB of McAulay ($e_{2,U}$ in Fig. \ref{06_varBBuplow_snr}) and thanks to the proposed interval variance approximations the MSEA is improved ($e_{1,U}$ and $e_{1,2,c}$ outperform $e_{1,1,c}$ in Fig. \ref{06_varBBapprox_snr}).
We have applied the MIE to non-oscillating ACRs. 
To the best of our knowledge this case is not considered before.


\section{
An AUB and an MSEA based on the interval probability
}\label{upperapprox_sec}

In this section we propose an AUB (Sec. \ref{upper_sec}) and an MSEA (Sec. \ref{approx_sec}), both based on the interval probability approximation $P^{(3)}_n$ in (\ref{proba3_eq}).


\subsection{An AUB}\label{upper_sec}

As $P^{(3)}_n$ approximates the probability that $\hat{\Theta}$ falls in $D_n$, the PDF of $\hat{\Theta}$ can be approximated by the limit of $P^{(3)}_n$ as $N$ (number of intervals) approaches infinity (so that the width of $D_n$ approaches zero). Accordingly we can write the approximate PDF, mean and MSE of $\hat{\Theta}$ as
\begin{eqnarray}
	p_M(\theta) &=& \lim_{N\rightarrow\infty}P^{(3)}_n = \frac{P(\theta,\Theta)}{\int_{\Theta_1}^{\Theta_2}P(\theta,\Theta)d\theta} \label{pdfM_eq}\\
	\mu_M &=& \int_{\Theta_1}^{\Theta_2}\theta p_M(\theta)d\theta \label{muM_eq}\\
	e_M &=& \int_{\Theta_1}^{\Theta_2}(\theta-\Theta)^2p_M(\theta)d\theta. \label{mseM_eq}
\end{eqnarray}
We will see in Sec. \ref{num_sec} that $e_M$ acts as an UB and also converges to a multiple of the CRLB.
In fact, $p_M(\theta)$ overestimates the true PDF of $\hat{\Theta}$ in the vicinity of $\Theta$ because it is obtained from $P^{(3)}_n$ which is in turn obtained from the interval probability UB $P^{(2)}_n$ in (\ref{proba2_eq}).


\subsection{An MSEA}\label{approx_sec}

To guarantee the convergence of the MSEA to the CRLB, we approximate the PDF of $\hat{\Theta}$ inside $D_0\approx[\Theta-\frac{\theta_1-\Theta}{2},\Theta+\frac{\theta_1-\Theta}{2})$ by $p_{0,N}(\theta)$ in (\ref{pdfnN_eq}) ($\Theta$ is the mean and $c(\Theta)$ is the MSE) and outside $D_0$ by 
$p'_M(\theta)=P(\theta,\Theta)\big/\int_{D_{\Theta}\setminus D_0}P(\theta,\Theta)d\theta$ 
(the corresponding mean and MSE are 
$\mu'_M=\int_{D_{\Theta}\setminus D_0}\theta p'_M(\theta)d\theta$ and 
$e'_M=\int_{D_{\Theta}\setminus D_0}(\theta-\Theta)^2p'_M(\theta)d\theta$),
and propose the following approximation:
\begin{eqnarray}
	p_{MN}(\theta) &=& (1-\tilde{P}_A)p_{0,N}(\theta) + \tilde{P}_Ap'_M(\theta) \label{pdfMN_eq}\\
	\mu_{MN} &=& (1-\tilde{P}_A)\Theta + \tilde{P}_A\mu'_M \label{muMN_eq}\\
	e_{MN} &=& (1-\tilde{P}_A)c(\Theta) + \tilde{P}_Ae'_M \label{mseMN_eq}
\end{eqnarray}
where $\tilde{P}_A=2P(\theta_1,\Theta)$ approximates the probability that $\hat{\Theta}$ falls outside $D_0$.
With oscillating ACRs, $\theta_1$ is the abscissa of the first local maximum after the global one; thus, $\theta_1\approx\Theta+\frac{1}{f_c(\Theta)}$.
With non-oscillating ACRs, the vicinity of the maximum is not clearly marked off; so, we empirically take $\theta_1=\Theta+\frac{\pi}{4\beta_s(\Theta)}$.

\smallskip

The first contribution in this section is the AUB $e_M$ which is very tight (as will be seen in Figs. \ref{06_varBBuplow_snr}  and \ref{06_varuplow_snr}) and also very easy to compute. The second one is the highly accurate MSEA $e_{MN}$ (as will be seen in Figs. \ref{06_varBBapprox_snr} and \ref{06_varapprox_snr}); to the best of our knowledge, this is the first approximation expressed as the sum of two terms when $X_{n_1},\cdots,X_{n_N}$ are correlated (see \cite{mcaulay3,Najjar2005,athley,Richmond2005,Richmond2006}).


\begin{figure}
  \centering
  \includegraphics[scale = 0.6]{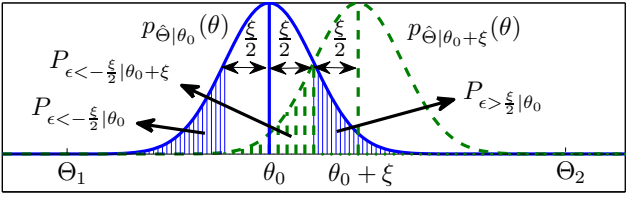}
  \caption{Decision problem with two equiprobable hypotheses: $H_1: \Theta=\theta_0$ and $H_2: \Theta=\theta_0+\xi$.} \label{05_2hypothesis_pic}
\end{figure}

\section{
ALBs
} \label{lower_sec}

In this section we derive an ALB based on the Taylor series expansion of the noise limited to second order (Sec. \ref{taylor_sec}) and a family of ALBs by employing the principle of binary detection which is first used by Ziv and Zakai \cite{ziv} to derive LBs for Bayesian parameters (Sec. \ref{zzb_sec}).


\subsection{
An ALB based on the second order Taylor series expansion of noise
} \label{taylor_sec} 

From (\ref{crosscorr2nd_eq}), the MLE of $\Theta$ can be approximated by:
\begin{equation} \label{mleC_eq}
	\hat{\Theta} = \argmax{\theta}{\{X_{s,r}(\theta)\}} \approx \hat{\Theta}_{C} = \Theta-\frac{\dot{w}_0}{\alpha\ddot{R}_0+\ddot{w}_0}
\end{equation}
where $\dot{w}_0/(\alpha\ddot{R}_0+\ddot{w}_0)$ is a ratio of two normal variables. Statistics of normal variable ratios are studied in \cite{Marsaglia1964,Marsaglia2006,Hinkley}.

\smallskip

Let $\operatorname{sign}(\xi)=1$ (resp. $-1$) for $\xi\geq0$ (resp. $\xi<0$),
$\delta^4(\theta)=E_{\ddot{s}}(\theta)/E_s$, $h=\operatorname{sign}(\nu_0)\sigma_{\dot{w}_0}\sqrt{1-\nu_0^2}$, $a_1=\nu_0\sigma_{\dot{w}_0}/\sigma_{\ddot{w}_0}$, $a_2=\sigma_{\ddot{w}_0}/h$, $a_3=\alpha\ddot{R}_0a_1/h$, $a_4=-\alpha\ddot{R}_0/\sigma_{\ddot{w}_0}=\sqrt{\rho}\beta^2(\Theta)/\delta^2(\Theta)$, $q(\xi)=(a_3\xi+a_4)/\sqrt{1+\xi^2}$.
We can show from \cite{Marsaglia2006} that $\hat{\Theta}_C$ in (\ref{mleC_eq}) is distributed as:
\begin{equation} \label{mleCpdflike_eq}
	\hat{\Theta}_C \sim \Theta+a_1+\frac{\chi}{a_2}
\end{equation}
where the PDF of $\chi$ is given by:
\begin{equation} \label{chipdf_eq}
	p_{\chi}(\xi) = \frac{e^{-\frac{a_3^2+a_4^2}{2}}}{\pi(1+\xi^2)}\Big\{1+\sqrt{2\pi}q(\xi)e^{\frac{q^2(\xi)}{2}}\Big(\frac{1}{2}-Q\big[q(\xi)\big]\Big)\Big\}.
\end{equation}
From (\ref{chipdf_eq}) we can approximate the PDF, mean, variance and MSE of $\hat{\Theta}_C$ by
\begin{eqnarray}
	p_C(\theta) &=& \operatorname{sign}(\nu_0)a_2p_{\chi}[a_2(\theta-\Theta-a_1)] \label{mleCpdf_eq}\\
	\mu_C &=& \int_{\Theta_1}^{\Theta_2}\theta p_C(\theta)d\theta \label{muC_eq}\\
	\sigma_C^2 &=& \int_{\Theta_1}^{\Theta_2}(\theta-\mu_C)^2p_C(\theta)d\theta \label{varC_eq}\\
	e_C &=& (\mu_C-\Theta)^2 + \sigma_C^2. \label{mseC_eq}
\end{eqnarray} 
Note that the moments $\int_{-\infty}^{\infty} \xi^i p_{\chi}(\xi) d\xi$, $i=1,2,\cdots$ (infinite domain) are infinite like with Cauchy distribution \cite{Marsaglia2006}.
We will see in Sec. \ref{num_sec} that $e_C$ behaves as an LB; 
this result can be expected from the approximation in (\ref{crosscorr2nd_eq}) where the expansion of the noise is limited to second order.


\subsection{
Binary detection based ALBs
} \label{zzb_sec}

Let $\tilde{\Theta}$ be an estimator of $\Theta$, $\epsilon|\theta=\tilde{\Theta}-\Theta$ the estimation error given $\Theta=\theta$, $p_{|\epsilon||\theta}(\xi)$ the PDF of $|\epsilon|$, and $P_{|\epsilon|>\xi|\theta}$ the probability that $|\epsilon|>\xi$. For $\Theta=\theta_0$, the MSE of $\tilde{\Theta}$ can be written as \cite{Cinlar}:
\begin{align}
	e|\theta_0 &= \int_0^{\epsilon_{\max}}\xi^2 p_{|\epsilon|\big|\theta_0}(\xi)d\xi
				= 2\int_0^{\epsilon_{\max}}\xi P_{|\epsilon|>\xi\big|\theta_0}d\xi \nonumber\\
						&- \{\xi^2P_{|\epsilon|>\xi\big|\theta_0}\}\big|_0^{\epsilon_{\max}}
				= \frac{1}{2}\int_{0}^{2\epsilon_{\max}}\xi P_{|\epsilon|>\frac{\xi}{2}\big|\theta_0}d\xi \label{localmse_eq}
\end{align}
where $\epsilon_{\max} = \max\{\Theta_2-\theta_0,\theta_0-\Theta_1\}$. 
By assuming $P_{\epsilon>\frac{\xi}{2}|\theta}$ and $P_{\epsilon<-\frac{\xi}{2}|\theta}$ constant $\forall\theta\in D_{\Theta}$, we can write \footnote{The obtained bounds are ``approximate" due to this assumption; the assumption is valid when $\theta$ is not very close to the extremities of $D_{\Theta}$.}:
\begin{align}
	P_{|\epsilon|>\frac{\xi}{2}|\theta_0} &= 2\left[\frac{1}{2}P_{\epsilon>\frac{\xi}{2}|\theta_0}+\frac{1}{2}P_{\epsilon<-\frac{\xi}{2}|\theta_0}\right] \label{localproba0_eq}\\
				& \approx 2\left\{\begin{array}{l}
					P_{\epsilon_1} = \frac{1}{2}P_{\epsilon>\frac{\xi}{2}|\theta_0-\xi} + \frac{1}{2}P_{\epsilon<-\frac{\xi}{2}|\theta_0} \\
					P_{\epsilon_2} = \frac{1}{2}P_{\epsilon>\frac{\xi}{2}|\theta_0} + \frac{1}{2}P_{\epsilon<-\frac{\xi}{2}|\theta_0+\xi}\end{array}\right. \nonumber\\
				&\geq 2\left\{\begin{array}{l}
					P_{\min}(\theta_0-\xi,\theta_0) \\
					P_{\min}(\theta_0,\theta_0+\xi)	\end{array}\right. \label{localproba_eq}
\end{align}
where $P_{\epsilon_1}$ and $P_{\epsilon_2}$ denote the probabilities of error of the nearest decision rule
\begin{equation} \label{Hnear_eq}
	\hat{H} = \Big\{{H_1 \atop H_2}\text{ if }|\tilde{\Theta}-\{\Theta|H_1\}|\lessgtr|\tilde{\Theta}-\{\Theta|H_2\}|
\end{equation}
of the two-hypothesis decision problems (the decision problem in (\ref{twohyp2_eq}) is illustrated in Fig. \ref{05_2hypothesis_pic}):
\begin{eqnarray}
	H &=& \left\{\begin{array}{ll}
				H_1: \Theta=\theta_0-\xi & P_{H_1}=0.5 \\
				H_2: \Theta=\theta_0 & P_{H_2}=0.5 \end{array}\right. \label{twohyp1_eq}\\
	H &=& \left\{\begin{array}{ll}
				H_1: \Theta=\theta_0 & P_{H_1}=0.5 \\
				H_2: \Theta=\theta_0+\xi & P_{H_2}=0.5 \end{array}\right. \label{twohyp2_eq}
\end{eqnarray}
and $P_{\min}(\theta_0-\xi,\theta_0)$ and $P_{\min}(\theta_0,\theta_0+\xi)$ the minimum probabilities of error obtained by the optimum decision rule based on the likelihood ratio test \cite[pp. 30]{wozencraft}:
\begin{equation} \label{optimal_eq}
	\hat{H} = \Big\{{H_1 \atop H_2} \text{ if } \Lambda(\Theta|H_1)-\Lambda(\Theta|H_2)\gtrless\ln\frac{P_{H_2}}{P_{H_1}}
\end{equation}
with $\Lambda(\theta)$ denoting the log-likelihood function in (\ref{loglikeli2_eq}). 
The probability of error of an arbitrary detector $\hat{H}$ is given by
\begin{equation}\label{Perr_eq}
	P_e = P_{H_1}P_{\hat{H}=H_2|H_1} + P_{H_2}P_{\hat{H}=H_1|H_2}.
\end{equation}
From (\ref{localmse_eq}) and (\ref{localproba_eq}) we obtain the following ALBs:
\begin{eqnarray}
	z_1 &=& \int_{0}^{\epsilon_1} \xi P_{\min}(\theta_0-\xi,\theta_0)d\xi \label{azzb1_eq}\\
	z_2 &=& \int_{0}^{\epsilon_2} \xi P_{\min}(\theta_0,\theta_0+\xi)d\xi \label{azzb2_eq}
\end{eqnarray}
%
%
where $\epsilon_1=\min\{\theta_0-\Theta_1,2(\Theta_2-\theta_0)\}$ and $\epsilon_2=\min\{\Theta_2-\theta_0,2(\theta_0-\Theta_1)\}$. The integration limits are set to $\epsilon_1$ and $\epsilon_2$ to make the two hypotheses in (\ref{twohyp1_eq}) and (\ref{twohyp2_eq}) fall inside $D_{\Theta}$.
As $P_{|\epsilon|>\frac{\xi}{2}|\theta_0}$ is a decreasing function, tighter bounds can be obtained by filling the valleys of $P_{\min}(\theta_0-\xi,\theta_0)$ and $P_{\min}(\theta_0,\theta_0+\xi)$ (as proposed by Bellini and Tartara in \cite{bellini}):
\begin{eqnarray}
	b_1 &=& \int_{0}^{\epsilon_1} \xi V\{P_{\min}(\theta_0-\xi,\theta_0)\}d\xi \label{abtb1_eq}\\
	b_2 &=& \int_{0}^{\epsilon_2} \xi V\{P_{\min}(\theta_0,\theta_0+\xi)\}d\xi \label{abtb2_eq}
\end{eqnarray}
where
$V\{f(\xi)\} = \max\{f(\zeta\geq\xi)\}$ denotes the valley-filling function.
When $P_{\min}(\theta,\theta')$ is a function of $\theta'-\theta$ (e.g, TOA estimation) we can write the bounds in (\ref{azzb1_eq})--(\ref{abtb2_eq}) as ($i=1,2$):
\begin{align}
	z_i &= \int_{0}^{\epsilon_i} \xi P_{\min}(\xi)d\xi \label{azzbi_eq}\\
	b_i &= \int_{0}^{\epsilon_i} \xi V\{P_{\min}(\xi)\}d\xi. \label{abtbi_eq}
\end{align}
If $\theta_0-\Theta_1>\Theta_2-\theta_0$, then $\epsilon_1>\epsilon_2$; hence, $z_1$ and $b_1$ become tighter than $z_2$ and $b_2$, respectively. 
From (\ref{loglikeli2_eq}), (\ref{Q_eq}), (\ref{optimal_eq}) and (\ref{Perr_eq}) we can write the minimum probability of error as
\begin{align}
	P_{\min}(\theta,\theta') &= 0.5\big[P_{\Lambda(\theta')>\Lambda(\theta)|\Theta=\theta}+P_{\Lambda(\theta)>\Lambda(\theta')|\Theta=\theta'}\big] \nonumber\\
				&= 0.5\big[P(\theta',\theta)|_{\Theta=\theta}+P(\theta,\theta')|_{\Theta=\theta'}\big] \nonumber\\
				&= Q\left(\sqrt{\frac{\rho}{2}[1-R(\theta,\theta')]}\right). \label{Pmin_toa_eq}
\end{align}

There are two main differences between our bounds (deterministic) and the Bayesian ones: i) with the former we integrate along the error only whereas with the latter we integrate along the error and the \textit{a priori} distribution of $\Theta$ (e.g, see (14) in \cite{bell}); ii) all hypotheses (e.g, $\Theta=\theta_0$ and $\Theta=\theta_0+\xi$ in \eqref{twohyp2_eq}) are possible in the Bayesian case thanks to the \textit{a priori} distribution whereas only one hypothesis ($\Theta=\theta_0$) is possible in the deterministic case. So in order to utilize the minimum probability of error we have approximated $P_{\epsilon<-\frac{\xi}{2}|\theta_0}$ in \eqref{localproba0_eq} by $P_{\epsilon<-\frac{\xi}{2}|\theta_0+\xi}$ (see Fig. \eqref{05_2hypothesis_pic}) .

\smallskip

In this section we have two main contributions. The first one is the ALB $e_C$ whereas the second one is the deterministic ZZLB family. These bounds can from now on be used as benchmarks in deterministic parameter estimation (like the CRLB) where it is not rigorous to use Bayesian bounds. Even though the derivation of $e_c$ was a bit complex, the final expression is now ready to be utilized.

\begin{figure}[t]
	\begin{center}
	\includegraphics[width=8.5cm]{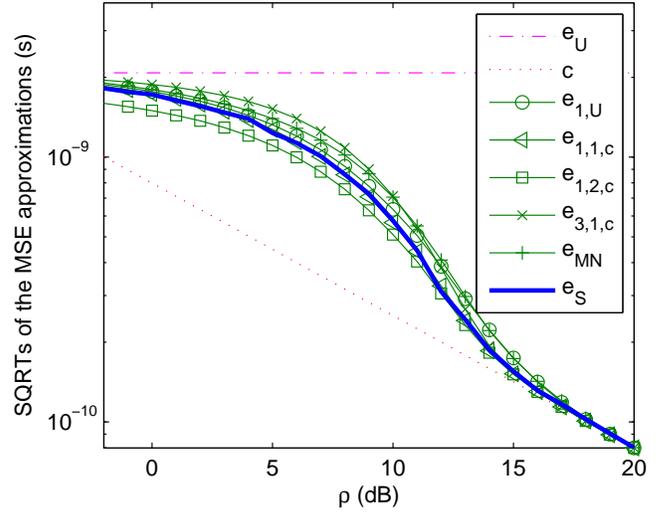}
	\caption{Baseband: SQRTs of the max. MSE $e_U$, the CRLB $c$, the MSEAs $e_{1,U}$, $e_{1,1,c}$, $e_{1,2,c}$, $e_{3,1,c}$ and $e_{MN}$, and the simulated MSE $e_S$, w.r.t. the SNR.}\label{06_varBBapprox_snr}
	\end{center}
\end{figure}


\section{Numerical results and discussion}\label{num_sec}

In this section we discuss some numerical results about the derived MSEAs, AUB, and ALBs.
We consider TOA estimation using baseband and passband pulses. Let $T_w=2$ ns, $f_c=6.85$ GHz, $\Theta=0$ and $D_{\Theta}=[-2,1.5]T_w$. With the baseband pulse we consider $9$ equal duration intervals.
Let
\begin{equation} \label{mseij_eq}
	e_{i,j,x} = P_0^{(i)}\sigma^2_{0,0} + \sum_{n=n_1,n\neq0}^{n_N}P_n^{(i)}\left[\left(\Theta-\mu_{n,j,x}\right)^2+\sigma^2_{n,j,x}\right]
\end{equation}
be the MSEA based on (\ref{mse_eq}) and using the interval probability approximation $P^{(i)}_n$ ($i\in\{1,2,3\}$, see (\ref{proba1_eq}), (\ref{proba2_eq}), (\ref{proba3_eq})) and interval mean and variance approximations $\mu_{n,j,x}$ and $\sigma^2_{n,j,x}$ ($(j,x)=U$ in (\ref{munU_eq}), (\ref{varnU_eq}), and $(j,x)\in\{1,2\}\times\{c,o\}$ in (\ref{mun1c_eq})--(\ref{varn2c_eq}), (\ref{mun1o_eq})--(\ref{varn2o_eq})).


\subsection{Baseband pulse}

\begin{figure}[t]
	\begin{center}
	\includegraphics[width=8.5cm]{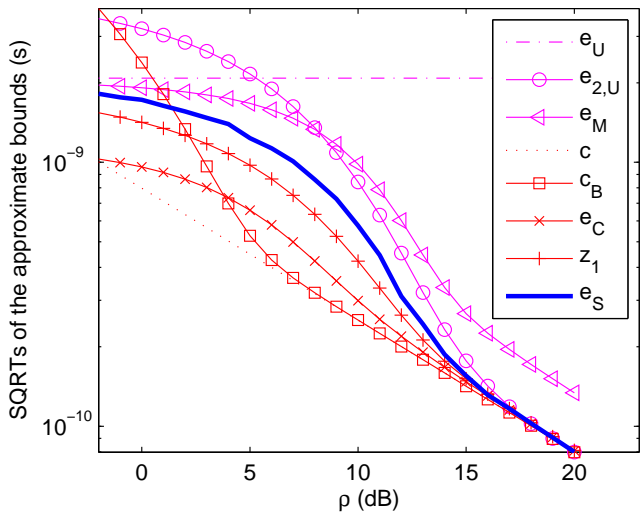}	
	\caption{Baseband: SQRTs of the max. MSE $e_U$, the AUBs $e_{2,U}$ and $e_M$, the CRLB $c$, the BLB $c_B$, the ALBs $e_C$ and $z_1$, and the simulated MSE $e_S$, w.r.t. the SNR.}\label{06_varBBuplow_snr}
	\end{center}
\end{figure}

Consider first the baseband pulse. 
In Fig. \ref{06_varBBapprox_snr} we show the SQRTs of the maximum MSE $e_U$ in (\ref{mseU_eq}), the CRLB $c$ in (\ref{crlb_eq}), five MSEAs: $e_{1,U}$, $e_{1,1,c}$, $e_{1,2,c}$, $e_{3,1,c}$ in (\ref{mseij_eq}) and $e_{MN}$ in (\ref{mseMN_eq}), and the MSE $e_S$ obtained by simulation based on 10000 trials, versus the SNR.
In Fig. \ref{06_varBBuplow_snr} we show the SQRTs of $e_U$, two AUBs: $e_{2,U}$ in (\ref{mseij_eq}) and $e_M$ in (\ref{mseM_eq}), $c$, the BLB $c_B$ in (\ref{barankin_eq}), two ALBs: $e_C$ in (\ref{mseC_eq}) and $z_1$ in (\ref{azzbi_eq}) (equal to $b_1$ in (\ref{abtbi_eq}) because a non-oscillating ACR), and $e_S$.

\smallskip

We can see from $e_S$ that, as cleared up in Sec. \ref{intro_sec}, the SNR axis can be divided into three regions: 1) the \textit{a priori} region where $e_U$ is achieved, 2) the threshold region and 3) the asymptotic region where $c$ is achieved. We define the \textit{a priori} and asymptotic thresholds by \cite{weiss2}:
\begin{eqnarray}
	\rho_{pr} &=& \rho \; : \; e(\rho)=\alpha_{pr}e_U \label{thpr_eq}\\
	\rho_{as} &=& \rho \; : \; e(\rho)=\alpha_{as}c. \label{thas_eq}
\end{eqnarray}
We take $\alpha_{pr}=0.5$ and $\alpha_{pr}=1.1$.
From $e_S$, we have $\rho_{pr}=4$ dB and $\rho_{as}=16$ dB.
Thresholds are defined in literature w.r.t. two magnitudes at least: i) the achieved MSE \cite{weiss2,zeira2,bell} like in our case (which is the most reliable because the main concern in estimation is to minimize the MSE) and ii) the probability of non-ambiguity \cite{Van1968,Richmond2005} (for simplicity reasons).

\smallskip

The MSEAs $e_{1,U}$, $e_{1,1,c}$, $e_{1,2,c}$, $e_{3,1,c}$ obtained from the MIE (Sec. \ref{MIE_sec}) are very accurate and follow $e_S$ closely; $e_{1,1,c}$ is more accurate than $e_{3,1,c}$ which slightly overestimates $e_S$ because $e_{1,1,c}$ uses the probability approximation $P_n^{(1)}$ in (\ref{proba1_eq}) that considers all testpoints during the computation of the probability, whereas $e_{3,1,c}$ uses the approximation $P_n^{(3)}$ in (\ref{proba3_eq}) based on the probability UB $P_n^{(2)}$ in (\ref{proba2_eq}) that only considers the $0$th and the $n$th testpoints;
$e_{1,1,c}$ is more accurate than $e_{1,U}$ which slightly overestimates $e_S$, and than $e_{1,2,c}$ which slightly underestimates it, because $e_{1,1,c}$ uses the variance approximation $\sigma^2_{n,1,c}$ in (\ref{varn1c_eq}) obtained from the first order Taylor series expansion of noise, whereas $e_{1,U}$ uses $\sigma^2_{n,U}$ in (\ref{varnU_eq}) assuming the MLE uniformly distributed in $D_n$ (overestimation of the noise), and $e_{1,2,c}$ uses $\sigma^2_{n,2,c}$ in (\ref{varn2c_eq}) neglecting the noise.
The MSEA $e_{MN}$ proposed in Sec. \ref{upper_sec} based on our probability approximation $P_n^{(3)}$ is very accurate as well.

\smallskip

The AUB $e_{2,U}$ proposed in \cite{mcaulay3} is very tight and converges to the asymptotic region simultaneously with $e_S$. However, it is less tight in the \textit{a priori} and threshold regions because it uses the probability UB $P_n^{(2)}$ which is not very tight in these regions (see Fig. \ref{06_Pn_snr_pic}). Moreover, $e_{2,U}\rightarrow\infty$ when $N\rightarrow\infty$.
The AUB $e_M$ (Sec. \ref{upper_sec}) is very tight. However, it converges to $2.68$ times the CRLB at high SNRs. This fact was discussed in Sec. \ref{upper_sec} and also solved in Sec. \ref{approx_sec} by proposing $e_{MN}$ (examined above). Nevertheless, $e_M$ can be used to compute the asymptotic threshold accurately because it converges to its own asymptotic regime simultaneously with $e_S$.

\smallskip

Both the BLB $c_B$ and the ALB $e_C$ (Sec. \ref{taylor_sec}) outperform the CRLB. Unlike the passband case considered below, $e_C$ outperforms the BLB.
The ALB $z_1$ (Sec. \ref{zzb_sec}) is very tight and converges to the CRLB simultaneously with $e_S$.

\begin{figure}[t]
	\begin{center}
	\includegraphics[width=8.5cm]{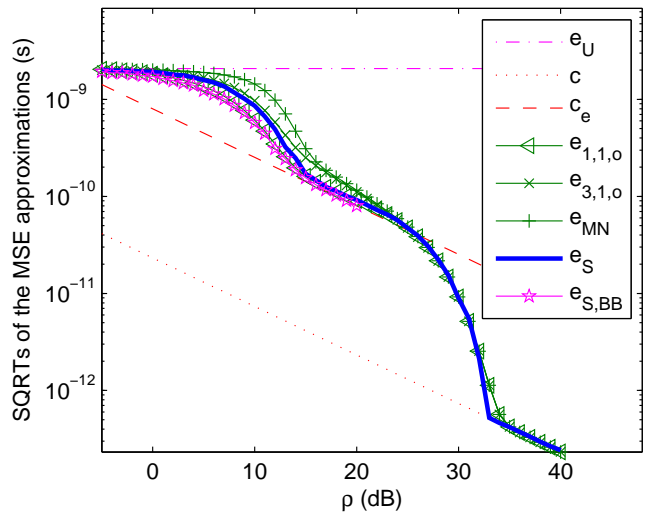}
	\caption{Passband: SQRTs of the max. MSE $e_U$, the CRLB $c$, the ECRLB $c_e$, the MSEAs $e_{1,1,o}$, $e_{3,1,o}$ and $e_{MN}$, and the simulated MSEs of the passabnd $e_S$ and baseband $e_{S,BB}$ pulses, w.r.t. the SNR.}\label{06_varapprox_snr}
	\end{center}
\end{figure}


\subsection{Passband pulse}

\begin{figure}[t]
	\begin{center}
	\includegraphics[width=8.5cm]{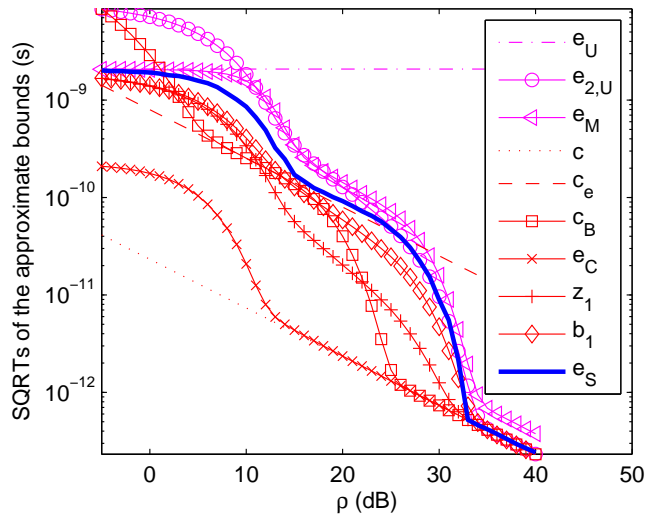}
	\caption{Passband: SQRTs of the max. MSE $e_U$, the AUBs $e_{2,U}$ and $e_M$, the CRLB $c$, the ECRLB $c_e$, the BLB $c_B$, the ALBs $e_C$, $z_1$ and $b_1$, and the simulated MSE $e_S$, w.r.t. the SNR.}\label{06_varuplow_snr}
	\end{center}
\end{figure}

Consider now the passband pulse. 
In Fig. \ref{06_varapprox_snr} we show the SQRTs of the maximum MSE $e_U$, the CRLB $c$, the ECRLB $c_e$ in (\ref{envcrlb_eq}) (equal to CRLB of the baseband pulse), three MSEAs: $e_{1,1,o}$ and $e_{3,1,o}$ in (\ref{mseij_eq}) and $e_{MN}$ in (\ref{mseMN_eq}), and the MSEs obtained by simulation for both the passband $e_S$ and the baseband $e_{S,BB}$ pulses.
In Fig. \ref{06_varuplow_snr} we show the SQRTs of $e_U$, two AUBs: $e_{2,U}$ in (\ref{mseij_eq}) and $e_M$ in (\ref{mseM_eq}), $c$, $c_e$, the BLB $c_B$, three ALBs: $e_C$ in (\ref{mseC_eq}), $z_1$ in (\ref{azzbi_eq}) and $b_1$ in (\ref{abtbi_eq}), and $e_S$.

\smallskip

By observing $e_S$, we identify five regions: 1) the \textit{a priori} region, 2) the \textit{a priori}-ambiguity transition region, 3) the ambiguity region where the ECRLB is achieved, 4) the ambiguity-asymptotic transition region and 5) the asymptotic region. 
We define the begin-ambiguity and end-ambiguity thresholds marking the ambiguity region by \cite{weiss2}
\begin{eqnarray}
	\rho_{am1} &=& \rho \; : \; e(\rho)=\alpha_{am1}c_e \label{tham1_eq}\\
	\rho_{am2} &=& \rho \; : \; e(\rho)=\alpha_{am2}c_e. \label{tham2_eq}
\end{eqnarray}
We take $\alpha_{am1}=2$ and $\alpha_{am2}=0.5$.
From $e_S$ we have $\rho_{pr}=7$ dB, $\rho_{am1}=15$ dB, $\rho_{am2}=28$ dB and $\rho_{as}=33$ dB.

\smallskip

The MSEAs $e_{1,1,o}$, $e_{3,1,o}$ (Sec. \ref{MIE_sec}) and $e_{MN}$ (Sec. \ref{approx_sec}) are highly accurate and follow $e_S$ closely.

\smallskip

The AUB $e_{2,U}$ \cite{mcaulay3} is very tight beyond the \textit{a priori} region.
The AUB $e_M$ (Sec. \ref{upper_sec}) is very tight. However, it converges to $1.75$ times the CRLB in the asymptotic region. 

\smallskip

The BLB $c_B$ detects the ambiguity and asymptotic regions much below the true ones; consequently, it does not determine accurately the thresholds ($\rho_{am1}=5$ dB, $\rho_{am2}=20$ dB and $\rho_{as}=26$ dB instead of 15, 28 and 33 dB).
The ALB $e_C$ (Sec. \ref{taylor_sec}) outperforms the CRLB, but is outperformed by the BLB (unlike the baseband case).
The ALB $z_1$ (Sec. \ref{zzb_sec}) is very tight, but $b_1$ (Sec. \ref{zzb_sec}) is tighter thanks to the valley-filling function. They both can calculate accurately the asymptotic threshold and to detect roughly the ambiguity region.

\smallskip

Let us compare the MSEs $e_{S,BB}$ and $e_S$ achieved by the baseband and passband pulses (Fig. \ref{06_varapprox_snr}).
Both pulses approximately achieve the same MSE below the end-ambiguity threshold of the passband pulse ($\rho_{am2}=28$ dB) and achieve the ECRLB between the begin-ambiguity and end-ambiguity thresholds.
The MSE achieved with the baseband pulse is slightly smaller than that achieved with the passband pulse because with the former the estimates spread in continuous manner along the ACR whereas with the latter they spread around the local maxima.
The asymptotic threshold of the baseband pulse (16 dB) is approximately equal to the begin-ambiguity threshold of the passband pulse (15 dB).
Above the end-ambiguity threshold, the MSE of the passband pulse rapidly converges to the CRLB while that of the baseband one remains equal to the ECRLB.
	
\smallskip

To summarize we can say that for a given nonlinear estimation problem with an oscillating ACR, the MSE achieved by the ACR below the end-ambiguity threshold is the same as that achieved by its envelope. Between the begin-ambiguity and end-ambiguity thresholds, the achieved MSE is equal to the ECRLB.
Above the latter threshold, the MSE achieved by the ACR converges to the CRLB whereas that achieved by its envelope remains equal to the ECRLB.


\section{Conclusion}
%
We have considered nonlinear estimation of scalar deterministic parameters and investigated the threshold and ambiguity phenomena.
The MIE is employed to approximate the statistics of the MLE. The obtained MSEAs are highly accurate and follow the true MSE closely.
A very tight AUB is proposed as well.
An ALB tighter than the CRLB is derived using the second order Taylor series expansion of noise.
The principle of binary detection is utilized to compute some ALBs which are very tight.


\appendices


\section{Curvatures of the ACR and of its envelope}\label{curvature_app}

In this appendix we prove (\ref{curvature_eq}). 
From (\ref{ft_eq}) and (\ref{cenv_eq}) we can write the FT of the complex envelope $e_{R_s}(\theta,\Theta)$ as
\begin{equation}\label{envft_eq}
	\mathcal{F}_{e_{R_s}}(f) = 2\mathcal{F}^+_{R_s}\left[f+f_c(\Theta)\right]
\end{equation}
where $x^+(f)=\left\{{x(f) \atop 0} {f>0 \atop f\leq0} \right.$.
Form (\ref{cenv_eq}) we can write
\begin{align}
	\ddot{R}_s(\theta,\Theta) & = \Re\Big\{e^{j2\pi(\theta-\Theta)f_c(\Theta)}\big[j4\pi f_c(\Theta)\dot{e}_{R_s}(\theta,\Theta) \nonumber\\
	& + \ddot{e}_{R_s}(\theta,\Theta) - 4\pi^2f_c^2(\Theta)e_{R_s}(\theta,\Theta)\big]\Big\} \label{curvature_app1_eq}
\end{align}
As from (\ref{cenv_eq}) $\Re\left\{e_{R_s}(\Theta,\Theta)\right\}=R_s(\Theta,\Theta)=E_s$, (\ref{curvature_app1_eq}) gives
\begin{align}
	\ddot{R}_s(\Theta,\Theta) &= \Re\big\{\ddot{e}_{R_s}(\Theta,\Theta)\big\}-4\pi^2f_c^2(\Theta)E_s \nonumber\\
	&+ 4\pi f_c(\Theta)\Re\big\{j\dot{e}_{R_s}(\Theta,\Theta)\big\}.	\label{curvature_app3_eq}
\end{align}
To prove (\ref{curvature_eq}) from (\ref{curvature_app3_eq}) we must prove that $\Re\{j\dot{e}_{R_s}(\Theta,\Theta)\}$ is null. 
Using (\ref{envft_eq}) and the inverse FT, we can write
\begin{align}
	&\dot{e}_{R_s}(\theta,\Theta) = \int_{-\infty}^{+\infty}j2\pi f\mathcal{F}_{e_{R_s}}(f)e^{j2\pi f(\theta-\Theta)}df \nonumber\\
	&= \int_{-\infty}^{+\infty}j4\pi f\mathcal{F}^+_{R_s}[f+f_c(\Theta)]e^{j2\pi f(\theta-\Theta)}df \nonumber\\
	&= \int_{-\infty}^{+\infty}j4\pi[f-f_c(\Theta)]\mathcal{F}^+_{R_s}(f)e^{j2\pi[f-f_c(\Theta)](\theta-\Theta)}df \nonumber\\
	&= \int_{0}^{+\infty}j4\pi[f-f_c(\Theta)]\mathcal{F}_{R_s}(f)e^{j2\pi[f-f_c(\Theta)](\theta-\Theta)}df \nonumber
\end{align}
so $\dot{e}_{R_s}(\Theta,\Theta) = \int_{0}^{+\infty}j4\pi[f-f_c(\Theta)]\mathcal{F}_{R_s}(f)df$.
%
%
%
Using (\ref{phic_eq}) and the last equation, $\Re\{j\dot{e}_{R_s}(\Theta,\Theta)\}$ becomes
\begin{equation}
	\Re\{j\dot{e}_{R_s}(\Theta,\Theta)\} = -\int_{0}^{+\infty}4\pi[f-f_c(\Theta)]\Re\{\mathcal{F}_{R_s}(f)\}df = 0. \nonumber
\end{equation}
Hence, (\ref{curvature_eq}) is proved.


\section*{Acknowledgment}
The authors would like to thank Prof. Alan Genz for his help in the probability numerical computation.


\footnotesize
\bibliographystyle{IEEEtranN}
\bibliography{IEEEabrv,all_chapters_ref_abbrev}


\end{document}